\documentclass[fleqn,usenatbib]{mnras}

\usepackage{prettyref}
\usepackage{float}
\usepackage{rotfloat}
\usepackage{amsmath}
\usepackage{amssymb}
\usepackage{graphicx}
\usepackage{color}
\usepackage{url}
\usepackage{comment}
\usepackage{natbib}

\usepackage[T1]{fontenc}
\usepackage{ae,aecompl}

\usepackage{tablefootnote}

\makeatletter
\providecommand{\tabularnewline}{\\}

\usepackage{indentfirst}
\usepackage{leftidx}

\graphicspath{{figures/}}
\makeatother

\begin{document}
\title[Scattering-Produced Polarization]{Scattering-Produced (Sub)millimeter Polarization in Inclined Disks: Optical Depth Effects, Near-Far Side Asymmetry, and Dust Settling}
\author[H. Yang et al.]{
Haifeng Yang$^{1}$\thanks{E-mail: hy4px@virginia.edu},
Zhi-Yun Li$^{1}$,
Leslie W. Looney$^{2}$,
Josep M. Girart$^{3}$,
\and
Ian W. Stephens$^{4}$
\\
$^{1}$Astronomy Department, University of Virginia, Charlottesville, VA 22904, USA\\
$^{2}$Department of Astronomy, University of Illinois at Urbana-Champaign, Urbana, IL 61801, USA\\
$^{3}$Institut de Ci\`{e}ncies de l'Espai (IEEC-CSIC), Campus UAB, Carrer de Can Magrans, s/n, E-08193 Cerdanyola del Vall\`{e}s, Catalonia, Spain\\
$^{4}$Harvard-Smithsonian Center for Astrophysics, 60 Garden Street, Cambridge, MA 02138, USA
}

\maketitle

\begin{abstract}
Disk polarization at (sub)millimeter wavelengths is being revolutionized by ALMA observationally, but its origin remains uncertain. Dust scattering was recently recognized as a potential contributor to polarization, although its basic properties have yet to be thoroughly explored. Here, we quantify the effects of optical depth on the scattering-induced polarization in inclined disks through a combination of analytical illustration, approximate semi-analytical modeling using formal solution to the radiative transfer equation, and Monte Carlo simulations. We find that the near-side of the disk is significantly brighter in polarized intensity than the far-side, provided that the disk is optically thick {\it and} that the scattering grains have yet to settle to the midplane. This asymmetry is the consequence of a simple geometric effect: the near-side of the disk surface is viewed more edge-on than the far-side. It is a robust signature that may be used to distinguish the scattering-induced polarization from that by other mechanisms, such as aligned grains. The asymmetry is weaker for a geometrically thinner dust disk. As such, it opens an exciting new window on dust settling. We find anecdotal evidence from dust continuum imaging of edge-on disks that large grains are not yet settled in the youngest (Class 0) disks, but become more so in older disks. This trend is corroborated by the polarization data in inclined disks showing that younger disks have more pronounced near-far side asymmetry and thus less grain settling. If confirmed, the trend would have far-reaching implications for grain evolution and, ultimately, the formation of planetesimals and planets.
\end{abstract}



\section{Introduction}
\label{sec:intro}


Polarized (sub)millimeter emission has been observed in an increasing number of
disks around young stellar objects. The original motivation for such observations is to
detect magnetic fields through linear dichroism of magnetically aligned grains
\citep{Cho2007,Bertrang2017,Andersson2015}; the fields are widely believed to play a crucial role in the
disk dynamics and evolution, through magnetorotational instability \citep{Balbus1991} and
magnetocentrifugal disk wind (\citealt{Blandford1982}; see \citealt{Turner2014} and
\citealt{Armitage2015} for recent reviews). The initial searches for disk polarization yielded
only upper limits \citep{Hughes2009,Hughes2013}. These were soon followed by detections in IRAS
16293B \citep{Rao2014} using the Submillimeter Array (SMA), in HL Tau \citep{Stephens2014}, L1527
\citep{Segura-Cox2015}, and Cepheus A HW2 \citep{Fernandez-Lopez2016} using the Combined
Array for Research in Millimeter-wave Astronomy (CARMA), and in NGC 1333 IRAS 4A at 8~mm and 1~cm
\citep{Cox2015,Liu2016} using the Karl G. Jansky Very Large Array (VLA). Most excitingly,
there are a large number of approved disk polarization programs using  Atacama Large
Millimeter/submillimeter Array (ALMA), with some results already published \citep{Kataoka2016b}
and several more in preparation (e.g., J. M. Girart, in prep). With its unique combination of high resolution and
sensitivity, ALMA ushers a new era of rapid growth in the observational study of disk
polarization.

Theoretical interpretation of the disk polarization at (sub)millimeter remains uncertain, however. The conventional interpretation is that the disk polarization comes from magnetically aligned non-spherical grains, as on the larger scales of molecular clouds and dense cores \citep{Andersson2015}. Initial calculations of disk polarization from aligned grains assume a purely toroidal magnetic field \citep{Cho2007,Bertrang2017}, as expected in a weakly magnetized disk \citep{Fromang2013}. Such a configuration would produce a polarization pattern that appears inconsistent with the pattern observed in HL Tau \citep{Stephens2014}. The apparent inconsistency led \cite{Yang2016a} to propose that the disk polarization in HL Tau comes from dust scattering, based on \cite{Kataoka2015} theory of scattering-induced millimeter polarization (see also \citealt{Kataoka2016a}). Specifically, they show that dust scattering in a disk inclined to the line of sight can naturally explain why the observed polarization vectors are roughly parallel to the minor axis and the distribution of polarized intensity is elongated along the major axis. This interpretation will be tested further in the near future with higher resolution ALMA observations at 3~mm and 0.87~mm, where the data have already been taken (PIs: A. Kataoka and I. Stephens, respectively).

There is evidence that scattering may also play a role in producing polarization in other sources, including Cepheus A HW2 \citep{Fernandez-Lopez2016} and HD 142527 \citep{Kataoka2016b}
at (sub) millimeter, NGC 1333 IRAS 4A at centimeter \citep{Cox2015, Yang2016b}, and AB Aur in mid-IR
\citep{Li2016}. If dust scattering really contributes significantly to the observed disk polarization at
(sub)millimeter and perhaps even centimeter wavelengths, the implication would be far-reaching: it would
provide direct evidence for grain growth, to sizes of order 100~$\mu$m or larger (\citealt{Kataoka2015}; see
also \citealt{Pohl2016}), which is the first step toward the formation of planetesimals and ultimately
planets. However, although the physics of dust scattering-induced polarization appears sound, whether it
indeed contributes significantly to the observed disk polarization is still not completely certain. For
example, in the case of HL Tau, \cite{Matsakos2016} shows that magnetically aligned grains may still be able
to explain the observation if the disk magnetic field is not dominated by a toroidal magnetic field, but
rather has a substantial radial component from, e.g, centrifugal disk winds. The situation is further
complicated by the possibility that large (non-spherical) grains may be aligned with the respect to the
direction of the anisotropy in the disk radiation field rather than the magnetic field \citep{Tazaki2017}.
These uncertainties highlight the need for finding additional distinguishing features of scattering-induced
disk polarization. Such features are required in order to use the disk  polarization to probe the grain
growth and/or magnetic field with confidence.


In this paper, we explore how the optical depth affects the disk polarization, focusing on dust scattering. The primary motivation comes from the fact that circumstellar disks tend to be optically thick at relatively short (sub)millimeter wavelengths. For example, the famous HL Tau disk, for which multi-wavelength polarization data have been taken by ALMA, is known to be optically thick at ALMA Band 7 (0.87~mm), especially within a radius of $\sim 50$~AU \citep{Carrasco-Gonzalez2016}. This is likely true for the disk around massive protostar HH 80-81 as well, for which polarization is also detected by ALMA ({J. M. Girart, in prep}). Such optically thick disks were not treated in our previous semi-analytical work on dust scattering, which was focused on the effects of disk inclination in the optically and geometrically thin limit. Here, we extend the treatment to include both a finite optical depth and a finite thickness for the emitting and scattering dust grains in the disk.
We find that, in an optically thick disk of a finite angular thickness in (vertical) dust distribution that is inclined significantly to the line of sight, the polarization pattern becomes asymmetric, with the near-side of the disk significantly brighter than the far-side in polarized intensity. As we will show later, the near-far side asymmetry is a simple consequence of the fact that, in an optically thick disk, the light detected by an observer comes mostly from the surface layer of the disk, which is inclined to the line of sight by a larger angle on the near-side than on the far-side (see Fig.~5 below for a cartoon illustration). This generic asymmetry, if detected, would not only add weight to the dust scattering interpretation of disk polarization, but also provide evidence that the large grains ($\sim 100\rm\, \mu m$) responsible for the scattering are not completely settled to the mid-plane.

The rest of the paper is organized as follows. We will start with an analytical illustration of the optical depth effects in a one-dimensional (1D) plane-parallel slab in \S~\ref{sec:analytical}, highlighting the difference between the polarization from scattering and from direct emission from aligned grains. The analytical results lay the foundation for interpreting the results obtained numerically under more complicated geometries. This is followed by radiative transfer calculations to quantify the optical depth effects on the scattering-produced polarization, especially the near-far side asymmetry in an inclined disk, in \S~\ref{sec:numerical}. We discuss the dependence of the near-far side asymmetry on the dust settling and observational evidence that large grains are less settled in younger disks in \S~\ref{sec:settling}, and the use of the near-far side asymmetry to distinguish the scattering-produced disk polarization from those from either magnetically or radiatively aligned grains in \S~\ref{sec:mechanism}. We conclude in \S~\ref{sec:conclusion} with our main results.

\section{Analytical Illustration of Optical Depth Effects}
\label{sec:analytical}

In order to illustrate how the optical depth affects the polarization analytically, we consider a slab of dust grains that is infinite in the $x$- and $y$-direction
but has a finite thickness in the $z$-direction. The grains are assumed to be isothermal and
uniformly distributed within the slab. Although our emphasis is on the scattering-produced polarization,
we will discuss in this section the polarization produced by direct emission from aligned (non-spherical) grains
as well, in order to contrast the optical depth effects in these two competing mechanisms.

\subsection{Polarization from scattering}
\label{subsec:semiinfinite}

For simplicity, we will consider Rayleigh scattering by spherical grains in this subsection. We denote the total optical depth of the slab along the $z$-direction by $\tau_{\rm max}$. The optical depth $\tau$ is defined as $d\tau = nC_{\rm ext}ds$, where $s$ is the distance along
the light path, $n$ the number density of the grains, and $C_{\rm ext}$ the extinction cross section. The radiative transfer
equation for the Stokes parameters vector $\mathbf{S}\equiv(I,Q,U,V)$ can be written as (see e.g. \citealt{TKS85}, Chapter 3):
\begin{equation}
	\frac{d\mathbf{S}}{ds} = -nC_{\rm ext} \mathbf{S} + nC_{\rm abs}
	\left(
	\begin{array}{c}
		B_\nu(T) \\ 0 \\ 0 \\ 0
	\end{array}
	\right) + n \int d\Omega \bar{M}(\tilde{\phi}) \bar{Z}(\Theta) \mathbf{S}_{\rm in}
	\label{eq:SRT}
\end{equation}
where $C_{\rm abs}$ is the absorption cross section, and
$\bar{Z}(\Theta)$ is the scattering phase matrix, which takes only one argument, the
scattering angle $\Theta$ between the incident and scattered light, because of the  assumed spherical shape for the grains.
The rotation matrix $\bar{M}(\tilde{\phi})$ transforms Stokes parameters from
the scattering plane coordinate---the coordinate system used for Rayleigh scattering---to
the lab frame. $\mathbf{S}_{\rm in}$ is the Stokes parameters vector for the incident light, which depends on the location of the scatterer and the direction of the light path in general.

The integral on the right hand side of Eq.\eqref{eq:SRT} can be treated as the source function for the scattered light. In this case,
scattering alone determines how the light is polarized since the direct emission from spherical grains is non-polarized, as indicated by the zeros in the second term. We will first consider an optically thick slab with $\tau_{\rm max}\gg 1$. If a scattering particle is located deep within the opaque slab, it would see a roughly isotropic incident radiation field and would produce little polarization in its scattered light. In contrast, a scattering particle located within a (vertical) optical depth of order unity or less ($\tau \lesssim 1$) of the surface sees an anisotropic incident radiation field, and has the potential of producing scattered light that is polarized.

\begin{figure}
	\centering
	\includegraphics[width=\columnwidth]{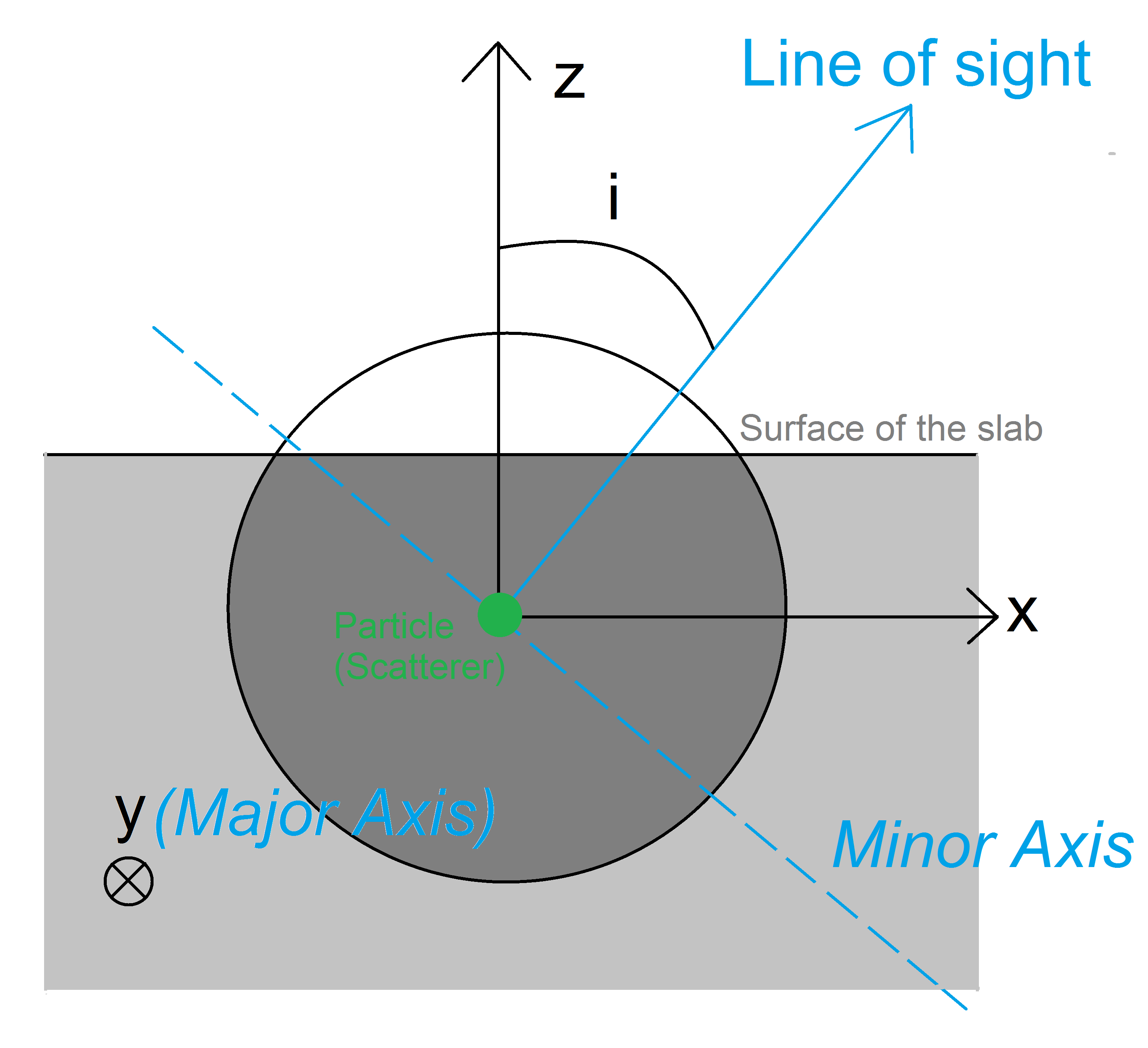}
	\caption{Coordinate system centered on a scattering particle located in the optically thin layer near the surface of the semi-infinite model. The dashed blue line denotes the direction of the local minor axis as viewed by the observer when the surface of the semi-infinite slab represents a local patch on the disk surface and the local major axis is along the y-axis.
	The circle indicates a sphere considered in the text around the scattering particle.}
	\label{fig:cartoon}
\end{figure}

To quantify the polarization from the light scattered by particles in the optically thin surface layer, we set up a coordinate system in the frame of the slab with $z$ directed upwards perpendicular to the slab (see Fig.~\ref{fig:cartoon}).  The line of sight direction lies in the $xz$-plane, making an inclination angle $i$ with respect to the $z$-axis.  As viewed by the observer, the $y$-axis of the slab (pointing into the page in Fig.~\ref{fig:cartoon}) lies in the plane of the sky.  If this slab represented a patch on the surface of an inclined disk, then the $y$-axis would correspond to the observed major axis; this will be important in \S~\ref{sec:numerical} below, and we will often refer to the $y$-axis as the ``major axis direction."  The axis perpendicular to the $y$-axis within the plane of the sky (represented by the blue dashed line in Fig.~\ref{fig:cartoon}) points along the intersection of the $xz$-plane with the plane of the sky, and it makes an angle $i$ with respect to the $x$-axis.  As viewed by the observer in the case of an inclined disk, this intersection would fall along the minor axis, and we will thus refer to it as the ``minor axis direction."

As we have previously made clear, the total polarization of this slab will be dominated by the radiation from the optically thin region.  We will first argue that the polarization direction of this light will be along the minor axis direction, as defined in the previous paragraph.  This can be demonstrated by considering the scattered radiation coming from a single scattering particle located at a vertical optical depth having $\tau < 1$.  Roughly speaking, the incident radiation as seen by this particle will reside within a sphere of radius corresponding to an optical depth of unity (represented by the circle in Fig.~\ref{fig:cartoon}), and if this sphere is centered on a particle within the optically thin region of the slab then a portion of the sphere will stick out above the upper surface.  The net polarization of the scattered light from this particle will then be determined by the net polarization of the incident radiation field it sees from this partially-filled sphere of material.

To determine what this incident polarization will be, let's first imagine that this sphere were completely filled with thermally-emitting dust grains.  If this were the case then the radiation field incident on the central scattering particle would be isotropic, and it would thus retain no net polarization after being scattered by that particle.  However, in the actual setup shown in Fig.~\ref{fig:cartoon}, a portion of the sphere (a ``cap," represented by the white unshaded segment) lies above the surface of the slab and does not contain any emitting dust grains.  The net polarization from this partially-filled sphere will then be equal to what we would expect from a completely filled sphere (i.e., zero) minus whatever would have been produced by the missing cap of material.  The radiation from the latter would primarily be directed along the $-z$ direction.  Because the polarization direction in Rayleigh scattering is perpendicular to both the incident radiation direction and the scattered radiation direction, the radiation from the missing material in the cap would be preferentially polarized along the $y$-axis (i.e., the major axis direction as seen in the plane of the sky) after being scattered by the particle at the center of the coordinate system.  Thus, after subtracting this component from the otherwise unpolarized scattered light that we would have from a completely filled sphere, the net result is a net polarization (for the setup shown in Fig. \ref{fig:cartoon}) that is along the minor axis direction.  This is a key feature that we will use to interpret the numerical results presented in \S~\ref{sec:numerical} below.

\begin{figure}
	\centering
	\includegraphics[width=\columnwidth]{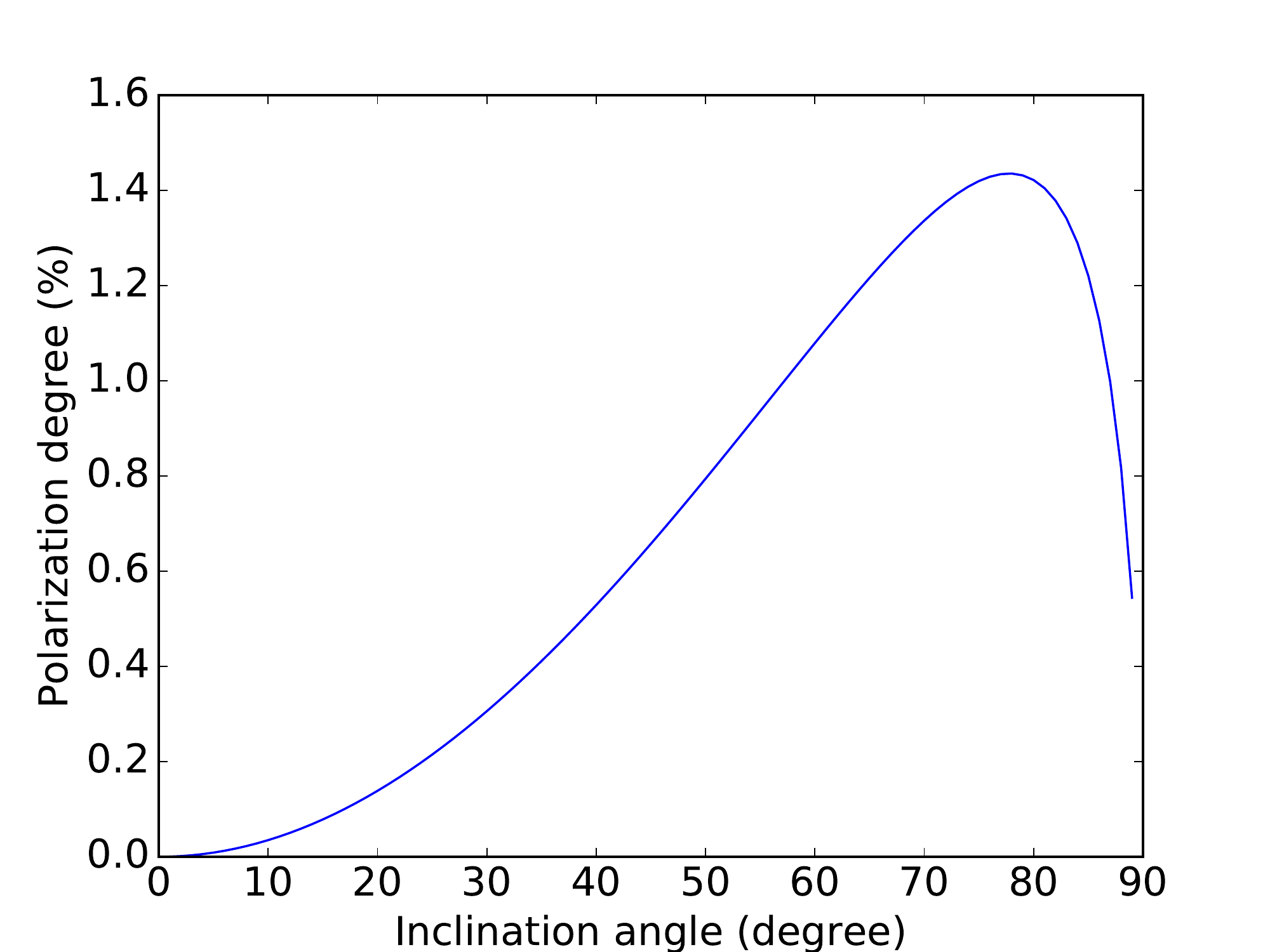}
	\caption{Maximum degree of polarization from scattering by spherical grains in a semi-infinite slab as a function of the slab inclination angle $i$ with respect to the line of sight ($i=0$ means face-on) under the single scattering approximation. The polarization peaks at $1.4\%$ for an inclination angle  $i=78^\circ$. The extinction opacity is assumed to be dominated by the scattering.}
	\label{fig:pincl}
\end{figure}

To calculate this degree of polarization along the minor axis direction in an inclined slab in a quantitative way, we need to solve the
integro-differential equation~(\ref{eq:SRT}) iteratively. For illustrative purposes, we will make the standard single scattering approximation, which is also assumed below in \S~\ref{subsec:fs} but will be checked in \S~\ref{subsec:mc} using Monte Carlo simulations that include multiple scattering. Under this approximation, the incident radiation $\mathbf{S}_{\rm in}$ on a  scattering particle located at a vertical optical depth $\tau$ below the surface is unpolarized and is given by the incident unpolarized intensity $I_{\rm in}$:
\begin{equation}
	\frac{I_{\rm in}(\theta, \tau)}{(C_{\rm abs}/C_{\rm ext})B_\nu(T)} = \left\{
	\begin{array}{ll}
		1-\exp\left[-\frac{\tau}{\cos(\theta)}\right] & 0\le\theta\le\frac{\pi}{2}\\
		1 & \frac{\pi}{2}\le\theta\le\pi
	\end{array}
	\right.,
	\label{eq:Iin}
\end{equation}
where $\theta$ is the polar angle measured from the $z$-axis.
All other components of Stokes parameters of the incident light are $0$. With
$\mathbf{S}_{\rm in}= (I_{\rm in}, 0, 0, 0)$, we can solve equation \eqref{eq:SRT} numerically.
For a semi-infinite slab ($\tau_{\rm max}\to\infty$) inclined by a representative angle $i=45^\circ$ to the line of sight, we have the Stokes $Q$:
\begin{equation}
	Q_{\rm \infty} = 0.012\times \frac{C_{\rm sca}}{C_{\rm ext}}
	\times \left(\frac{C_{\rm abs}}{C_{\rm ext}}\right)B_\nu(T),
	\label{eq:Qsinf}
\end{equation}
where $Q$ is positive along the minor axis direction in the plane of the sky, and
\begin{equation}
    I_{\rm \infty} = \left(1 + 0.81 \times \frac{C_{\rm sca}}{C_{\rm ext}}\right)
    \frac{C_{\rm abs}}{C_{\rm ext}}B_\nu(T).
    \label{eq:Isinf}
\end{equation}
In the total intensity shown in Eq.\eqref{eq:Isinf}, the first term comes from direct thermal emission.
It differs from the Planck function $B_\nu(T)$ because of the scattering in the material.
The second term comes from scattering and has the same dependence as $Q_{\infty}$ shown in Eq.\eqref{eq:Qsinf}.
The Stokes $U$ and $V$ are expected to be zero.
If we take a ratio of $Q_\infty$ and the second term of $I_\infty$, we get
$1.47\%$. This is the polarization degree of the purely scattered light. It is much smaller than that in the geometrically and optically thin disk cases, which can be as
high as $\sim 20\%$ \citep{Yang2016a}. The actual polarization degree (relative to the total intensity, not just the scattered intensity) can be expressed as:
\begin{equation}
    p_{\rm sca,\infty} = \frac{0.012C_{\rm sca}}
    {C_{\rm ext}+0.81C_{\rm sca}}.
\end{equation}
We can see that, for a given inclination angle $i$, the polarization degree depends only on the ratio of $C_{\rm ext}/C_{\rm sca}$, and that it reaches a maximum value when $C_{\rm ext}\rightarrow C_{\rm sca}$, i.e., the extinction is dominated by scattering rather than absorption. In this case, we have $p_{\rm sca,\infty}=0.66\%$ for $i=45^\circ$. In Fig.~\ref{fig:pincl}, we plot the  maximum degree of polarization for a semi-infinite slab as a function of the inclination angle under the single scattering approximation.
It is clear that the degree of polarization increases with the inclination angle $i$ except when the
line of sight becomes so inclined that it is nearly parallel to the slab. The increase comes about because
the light coming from the $y$-axis direction in the slab is always scattered by $90^\circ$ by a  particle
located in the optically thin surface layer (where $\tau < 1$) into the line of sight (and it is thus full
polarized for Rayleigh scattering) while that coming from the $x$-axis direction is scattered by an angle that
becomes closer to 0 or $180^\circ$ (and thus less polarized) as the inclination angle $i$ increases. This
is the same argument that we used in \cite{Yang2016a} to explain the increase of polarization degree
with inclination angle in a geometrically and optically thin disk. The difference is that this trend
continues all the way to  $i=90^\circ$ in \cite{Yang2016a} but not here, because of optical depth
effects: the scattered light takes an increasingly longer path out of the slab and thus  gets more
attenuated as the line of sight becomes more parallel to the surface of the slab.  

\begin{figure}
	\centering
	\includegraphics[width=\columnwidth]{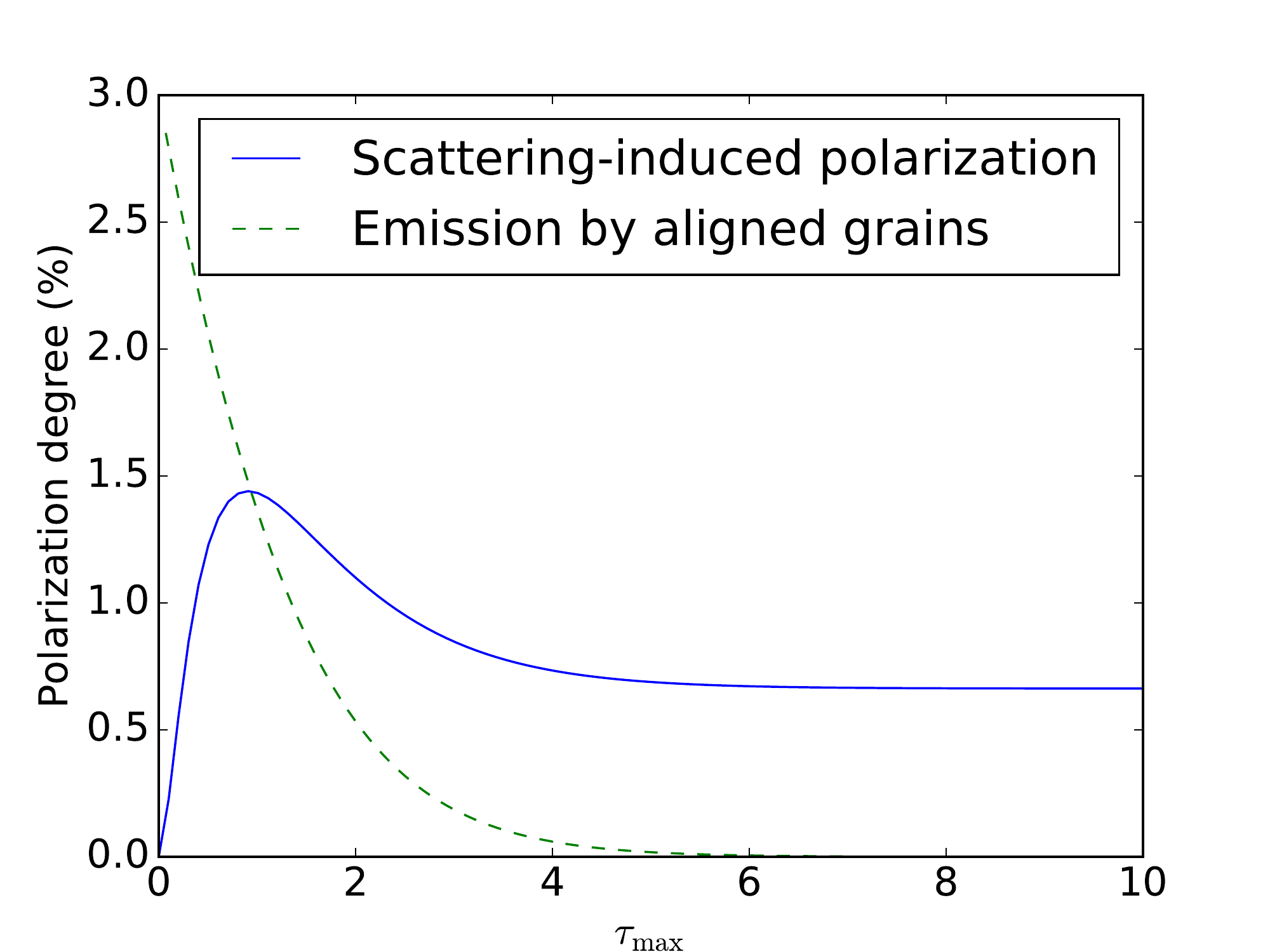}
	\caption{Variation of the degree of polarization from dust scattering (blue line) and direct emission from aligned grains (dashed green line) with the total optical depth $\tau_{\rm max}$ of an isothermal slab with an inclination angle of $45^\circ$. The polarization degree from aligned grains is assumed to be $3\%$ at small optical depth. Note the contrasting behaviors of the two curves at both small and large optical depth limits. See text for discussion.}
	\label{fig:ptau}
\end{figure}

We can generalize equation \eqref{eq:Iin} to the case of a finite slab
of (vertical) optical depth $\tau_{\rm max}$ to:
\begin{equation}
	\frac{I_{\rm in}(\theta, \phi, \tau, \tau_{\rm max})}{(C_{\rm abs}/C_{\rm ext})B_\nu(T)} = \left\{
	\begin{array}{ll}
		1-\exp\left[-\frac{\tau}{\cos(\theta)}\right] & 0\le\theta\le\frac{\pi}{2}\\
		1-\exp\left[-\frac{\tau_{\rm max}-\tau}{|\cos(\theta)|}\right] & \frac{\pi}{2}\le\theta\le\pi\\
	\end{array}
	\right..
	\label{eq:Iin_new}
\end{equation}
With the incident radiation field specified, it is again straightforward to integrate the radiative
transfer equation \eqref{eq:SRT} to determine the degree of polarization viewed from any inclination
angle $i$. As an example, we show in Fig. \ref{fig:ptau} the degree of polarization as a function
of the slab optical depth $\tau_{\rm max}$ for a representative angle $i=45^\circ$.
For an optically thin slab with $\tau_{\rm max} \lesssim 1$, the degree of polarization increases with the
optical depth of the slab for a fixed inclination angle. This is expected because, as already pointed
out in \cite{Yang2016a}, a higher slab optical depth $\tau_{\rm max}$ means more particles emitting
more photons to be scattered {\it and} more scattering particles, the combination of which yields a larger increase in the
intensity of the scattered light compared to that of the thermally emitted (non-polarized) light.
In the opposite limit of a highly opaque slab ($\tau_{\rm max}\gg 1$),  the polarization degree
asymptotes to a constant value, which is $0.66\%$ for the inclination angle adopted here ($i=45^\circ$). Such opaque
slabs are effectively semi-infinite, with a polarization degree approaching that shown in Fig.~\ref{fig:pincl}. Note that the polarization degree
peaks for a translucent slab with $\tau_{\rm max}\approx 1$, at a value of $1.44\%$, which
is significantly above the asymptotic value. One can understand this result qualitatively with the help of
a sphere of unit optical depth as before: most of the polarization observed outside the slab (from
above, say) comes from the scattering particles located in the optically thin layer below the
surface, and the unit optical depth sphere centered on such a particle would stick out of both the
top and bottom surfaces of the translucent slab, leading to two missing caps (instead of just one)
and thus a more anisotropic incident radiation field and a more polarized scattered light compared
to the semi-infinite case.

\subsection{Polarization from direct emission}


So far we have only considered spherical dust grains. Light emitted by such dust grains is
not polarized, which is also true for randomly oriented non-spherical dust grains.
Direct emission from aligned non-spherical grains produces polarization that depends on the
optical depth very differently from that from scattering.
Non-spherical grains tend to rotate with their shortest principle axis parallel to the
external ``alignment axis'', which is usually taken to be the magnetic field direction, although other possibilities exist (e.g., \citealt{Tazaki2017}).
Because of their linear dichroism, such grains have different optical depths for the light
polarized parallel and perpendicular to the alignment axis, which are denoted by
$\tau_\parallel$ and $\tau_\perp$ respectively. In the simplest case of uniform grain
alignment orientation and isothermal condition,
the expected polarization is \citep{Andersson2015, Hildebrand2000}:
\begin{equation}
	p_{\rm emit} = -\frac{e^{-\tau}\sinh(p_0\tau)}{1-e^{-\tau}\cosh(p_0\tau)},
	\label{eq:peandersson}
\end{equation}
where $\tau=(\tau_{\perp}+\tau_{\parallel})/2$ is the (averaged) optical depth
along line of sight, and the sign of the polarization degree is defined such that positive
means polarization parallel to the alignment axis.
Note that $p_{\rm emit}$ is always negative, which means that the polarization
from direct emission in this isothermal case is, as expected, perpendicular to the alignment axis. The parameter $p_0$ is defined as
\begin{equation}
    p_0= \frac{\tau_{\perp}-\tau_{\parallel}}{\tau_{\perp}+\tau_{\parallel} },
\end{equation}
whose value depends on the intrinsic properties of the grain.
In the slab model of a finite thickness that we discussed above,
the total optical depth along line of sight is
$\tau_{\rm max}/\cos(i)$, where $i$ is the inclination angle. So we have:
\begin{equation}
	p_{\rm emit} = -\frac{e^{-\tau_{\rm max}/\cos(i)}\sinh(p_0\tau_{\rm max}/\cos(i))}{1-e^{-\tau_{\rm max}/\cos(i)}\cosh(p_0\tau_{\rm max}/\cos(i))},
	\label{eq:pemit}
\end{equation}
It is clear that the polarization from direct emission is $p_0$ when $\tau_{\rm max} \ll 1$ and it decreases monotonically as the optical depth increases, approaching
$0$ exponentially for $\tau_{\rm max} \gg 1$. This behavior is very different from the polarization produced by scattering, which increases with the optical depth $\tau_{\rm max}$ of the slab initially, peaking around $\tau_{\rm max} \sim 1$, before asymptoting to a finite value. 
The polarization fraction for emission from aligned grains in the optically thin limit, $p_0$, depends on many factors, including grain shape, composition and alignment efficiency. It can be quite high on parsec scales or larger (up to $\sim 20\%$; \citealt{PlanckXIX2014}), although the (sub)millimeter polarization fraction detected on the scale of protoplanetary disks is typically on the order of a few percent or less (e.g., Stephens et al. 2014). For illustration purposes, we adopt $p_0=3\%$ in Fig.~\ref{fig:ptau},
where the polarization from direct emission by an isothermal slab is plotted as a function of $\tau_{\rm max}$ together with that produced by scattering 
for the same slab inclination of $i=45^\circ$.  
Clearly, the polarization is more likely dominated by scattering than direct emission if the optical depth of the slab is large {\it and}  the temperature is spatially constant.\footnote{
The optical depth at which the polarization fractions from emission by aligned grains and scattering become equal depends on the choice of $p_0$. For example, its value would change from $\tau_{\max}\approx 1$ to about 4 when $p_0$ is increased from $3\%$ to $20\%$.}

For an optically thick slab with a vertical temperature gradient, the above result no longer holds. As an illustration, we adopt a temperature structure
$T(\zeta) = T_0 + a \zeta$, where both $T_0$ and $a$ are constant and $\zeta$ is the distance along line of sight with $\zeta=0$ corresponding to the surface of
the slab facing the observer. For a semi-infinite slab (with $\zeta$ going from 0 to
$\infty$), the standard Eddington-Barbier relation in radiative transfer yields
the intensity at the surface along a given direction, which is equal to the source function
at an optical depth of $1$ along that direction. However, since the opacity is different for
light polarized parallel and perpendicular to the axis of grain alignment, differently
polarized light reaches optical depth of $1$ at different physical depths. In this non-isothermal case, the polarization degree is given by:
\begin{equation}
	p_{\rm emit}' = \frac{\frac{1}{2}a(\alpha_\perp-\alpha_\parallel)}{T_0\alpha_\perp\alpha_\parallel +
		\frac{1}{2}a(\alpha_\perp+\alpha_\parallel)}.
	\label{eq:pemitp}
\end{equation}
where we've assumed Rayleigh-Jeans limit for the thermal emission and
$\alpha_{\parallel(\perp)} \equiv \tau_{\parallel(\perp)}/\zeta$ is the absorption coefficient for light
with polarization parallel (perpendicular) to the alignment axis. In general, dust
grains are aligned with their long axis perpendicular to the alignment axis, so that $\alpha_\perp>\alpha_\parallel$.
As a result, the above expression gives a positive polarization degree, which means that the
polarization is parallel to the alignment axis for a positive temperature gradient
$a$. This is the opposite of the isothermal case, where the polarization is perpendicular
to the alignment axis (see Eq. \eqref{eq:peandersson}). In the case of grains aligned by a magnetic field,
the alignment axis will be the field direction. Although the general expectation is that the thermal
dust emission is polarized perpendicular to the field direction, Equation~(\ref{eq:pemitp}) demonstrates that
this may not be true in the presence of a temperature gradient along the line of sight. This simple exercise drives home the point that, in optically thick regions, the interpretation of polarization from aligned grains is not as straightforward as in the familiar optically thin limit, and can depend strongly on temperature distribution. We will postpone a more detailed exploration of this important point to a future investigation. 

\section{Near-far side asymmetry in the polarization of inclined disks}
\label{sec:numerical}

In this section, we aim to quantify the effects of optical depth on the scattering-induced
polarization in an inclined disk of young stellar object, focusing on the difference
between the near and far side of the disk. For simplicity, we will assume
spherical grains as in \S~\ref{subsec:semiinfinite}, and postpone a treatment of scattering by
aligned non-spherical grains to a future investigation. We will solve the
transfer equation (1) for the polarized light in two
complementary ways: using formal solution under the approximation of single
scattering (\S~\ref{subsec:fs}) and through Monte Carlo simulations
(\S~\ref{subsec:mc}). The former is conceptually straightforward and mathematically
simple, involving only direct integration along straight lines. It is well
suited for illustrating the basic effects of optical depth. The latter is
more general, with multiple scattering taken into account self-consistently, but produces noisier results that are harder to interpret physically.

\subsection{Formal solution under single scattering approximation}
\label{subsec:fs}

In the presence of scattering, the radiative transfer equation (1) is an
integro-differential equation that is difficult to solve in general. The
solution is simplified by the single scattering approximation, where the
incident radiation ${\bf S}_{\rm in}$ to be
scattered by dust grains at any location is assumed to come solely from direct
thermal emission. Since the dust grains are assumed to be spherical
and thus emit only unpolarized light, $\mathbf{S}_{\rm in}$ can be written as
$(I_{\rm in}, 0, 0, 0)$, with $I_{\rm in}$ determined by:
\begin{equation}
	\frac{dI_{\rm in}}{ds'} = -nC_{\rm ext} I_{\rm in} + nC_{\rm abs} B_\nu(T),
	\label{eq:IRT}
\end{equation}
which has the following formal solution:
\begin{equation}
	I_{\rm in} = \int_0^\infty e^{-\int_0^{s'} n(s'')C_{\rm ext}ds''} n(s') C_{\rm abs} B_{\nu}(T(s')) ds'.
	\label{eq:formal_sol}
\end{equation}
In the above expression, we have explicitly written the dependence of number density $n$ and
temperature $T$ on the location, described by the distance $s^\prime$ from the scatterer along a given direction ${\bf n}^\prime$. The cross
sections $C_{\rm ext}$ and $C_{\rm abs}$ are taken to be constant for simplicity, although they could vary spatially in general, due to, e.g., variation in grain size distribution. Such additional complications will be explored in future investigations.

Once the incident radiation ${\bf S}_{\rm in}$ is determined, we can compute through straightforward integration the Stokes parameters for the scattered light ${\bf S}_{\rm sca}$ (the third term on the right hand side of equation [1]) which, together with the (non-polarized) thermal dust emission, serves as the source for the formal solution that determines the Stokes parameters along any line of sight through the disk.


\subsubsection{Disk model}

For illustration purposes, we will adopt the disk model used by \cite{Kwon2011} to fit the millimeter observations of the HL Tau disk. It is a standard viscous accretion disk model with a density profile for dust grains only (assuming a dust-to-mass ratio of 100):
\begin{equation}
	\rho(R, z) = \rho_0 \left( \frac{R}{R_c} \right)^{-p}
	\exp\left[ -\left(\frac{R}{R_c}\right)^{3.5-p-q/2} \right]
	\exp\left[ -\left(\frac{z}{H(R)}\right)^2 \right],
	\label{eq:HLTaurho}
\end{equation}
where $R$ and $z$ are the coordinates for a cylindrical coordinate system, $R_c$ a characteristic radius of the disk (dust) density distribution, and $H(R)$ the (dust) scale height at a radius $R$, determined by hydrostatic equilibrium in the vertical direction in the case of no dust settling (the effects of dust settling will be discussed in \S~\ref{sec:settling} below). The scale height scales with radius as $H(R) = H_0(R/R_c)^{1.5-q/2}$. 

The $q$ parameter is the temperature power law index $T\propto R^{-q}$.
To account for the surface heating by stellar irradiation approximately, a two-component temperature distribution is adopted, with $T(R, z) = W T_m(R)+(1-W)T_s(r)$, where
$T_m(R)=T_0(R_0/R)^q$ and $T_s(r)=T_{s0}(r_{s0}/r)^q$ are the mid-plane and
surface temperature distribution, respectively, $r=\sqrt{R^2+z^2}$ is the spherical radius, and $W={\rm
exp}[-(z/3H(R))^2]$ is chosen to mimic the temperature profile computed self-consistently. The values for the
parameters that \cite{Kwon2011} found to be the best fit to the HL Tau disk data are listed in Table \ref{tab:par}. We leave the
density scale $\rho_0$ as a free parameter, in order to explore the effects of the disk column density (and
thus optical depth for a given grain size distribution). We will consider a range of values for $\rho_0$,
including an extreme case with $\rho_0=1.124\times 10^{-14} \rm\, g/cm^3$, corresponding to a disk mass of
about $1$ solar mass (motivated by the ALMA polarimetric observations of the HH80-81/IRAS 18162$-$2048 massive protostar, J.M. Girart, in
preparation, see discussions in \S~\ref{subsec:asymmetry}), where the optical depth effects are most apparent.

\begin{table}
	\centering
	\begin{tabular}{|c|c|c|c|}
		\hline
		$R_c$ & $79$ AU\ \ \ \ \ \  & \ \ \ \ \ \ \ $p$ & $1.064$ \\
		\hline
		$q$ & $0.43$ \ \ \ \ \ \  & \ \ \ \ \ \ \ $H_0$ & $16.8$ AU \\
		\hline
		$T_0$ & $70$ K \ \ \ \ \ \  & \ \ \ \ \ \ \ $R_0$ & $10$ AU \\
		\hline
		$T_{s0}$ & $400$ K \ \ \ \ \ \  & \ \ \ \ \ \ \ $r_{s0}$ & $3$ AU \\
		\hline
	\end{tabular}
	\caption{Parameters for the disk model.}
	\label{tab:par}
\end{table}

\begin{table}
    \centering
    \begin{tabular}{|c|c|c|c|}
        \hline
       Model name  & $a$ ($\mu m$) & $\rho_0$ ($\rm g/cm^3$) & $H_0$ (AU) \\
        \hline
        A & $100$ & $1.124\times 10^{-16}$ & 16.8 \\
        B & $100$ & $1.124\times 10^{-15}$ & 16.8 \\
        C & $100$ & $1.124\times 10^{-14}$ & 16.8 \\
        D & $37.5$ & $1.124\times 10^{-14}$ & 16.8 \\
        E & $10$ & $1.124\times 10^{-14}$ & 16.8 \\
        F & $100$ & $1.124\times 10^{-13}$ & 1.68 \\
        \hline
    \end{tabular}
    \caption{Different models used in this paper.}
    \label{tab:model}
\end{table}

For grain properties, we adopt the spherical dust grain model used in \cite{Kataoka2015}
and \cite{Yang2016a}. It is a composite grain model with abundances consistent with
\cite{Pollack1994}, which are composed of $8\%$ silicate, $62\%$ water ice and $30\%$ organic by volume.
For illustration purposes, we will first consider single-sized spherical dust grains\footnote{We only consider single-sized grains in this work. In the limit of Rayleigh scattering under consideration, grains of a range of sizes can be represented by single-sized grains of a certain (equivalent) radius. }
with radius $100\rm\,\mu m$ to maximize the effects of scattering at 1~mm; smaller grains that scatter millimeter light less efficiently are considered in \S~\ref{subsec:mc}. At $1$~mm wavelength, the scattering and absorption opacities are $\kappa_{\rm sca}=6.35\rm\, cm^2/g$ and
$\kappa_{\rm abs}=0.738\rm\, cm^2/g$, respectively, based on Mie theory \citep{BH83}.

\subsubsection{Results}
\label{subsubsec:results}

\begin{figure*}
	\centering
	\includegraphics[width=\textwidth]{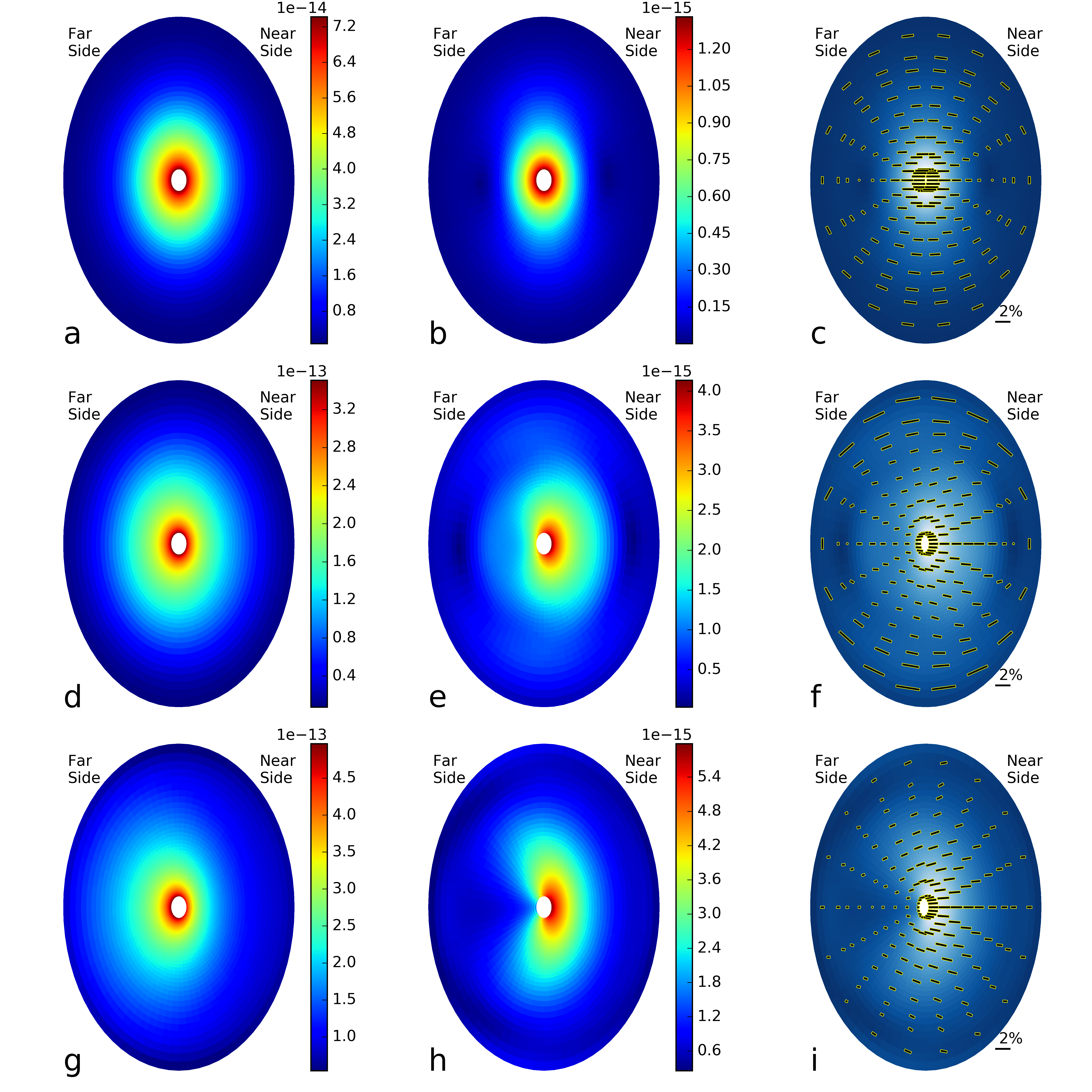}
	\caption{Distributions of the total intensity (left panels), polarized intensity (middle) and polarization vectors (right panels, with length proportional to the polarization degree) for three disk models with increasing density scale $\rho_0=1.13\times 10^{-16}$ (Model A; top panels), $1.13\times
	10^{-15}$ (Model B; middle), and $1.13\times 10^{-14}$~g~cm$^{-3}$ (Model C; bottom).
	Intensities are in unit of
    $\rm erg\cdot s^{-1}\cdot sr^{-1}\cdot cm^{-2}\cdot Hz^{-1}$. }
	\label{fig:image}
\end{figure*}

In this subsection, we present and discuss the results of integrating the formal solution
numerically, focusing on the effects of the optical depth in a disk inclined to the line of sight by $45^\circ$, similar to the value obtained for the HL Tau disk \citep{ALMA2015,Kwon2011}. To make connection with previous
(semi-analytical) work, we start with an optically thin case, with the density scale $\rho_0$
set to $1.13\times 10^{-16}$~g~cm$^{-3}$ (Model A in Table~\ref{tab:model}). This corresponds to a characteristic absorption
optical depth vertically through the disk at the characteristic radius $R_c=79$~AU of
$\tau_{\rm c,abs}=0.0136$ at 1 mm wavelength; the scattering optical depth is larger
($\tau_{\rm c,sca}=0.117$), but still much less than unity.  In this case, the total intensity, shown in Fig.~\ref{fig:image}a, appears symmetric between the near- and far-side (to the left and right of the major axis respectively), as expected for an optically thin disk. The polarization
pattern, shown in Fig.~\ref{fig:image}b and \ref{fig:image}c, is very similar to that obtained by \cite{Yang2016a} semi-analytically based on the simplification that the disk is both optically and
geometrically thin (see their Fig.~7). Specifically, the distribution of the polarized
intensity is elongated along the major axis and the polarization vectors in the central
region are oriented roughly along the minor axis. As discussed in depth in \cite{Yang2016a}, both of these features are the consequences of a simple geometric effect: the
incident light along the major axis is scattered by $90^\circ$ into the line of sight and is
thus maximally polarized. The polarization degree (about $1.5\%$ here) is also quantitatively
consistent with that in \cite{Yang2016a} (slightly bigger than $1\%$) for the following
reason. The grain size $a$  adopted here is about 3 times larger ($100\rm\,\mu m$ vs $36\rm\,\mu m$), which makes the
scattering opacity ($\propto a^3$ for Rayleigh scattering) and thus the polarization degree a factor of about $3^3=27$ larger. On the other hand, the density scale $\rho_0$ for this model is about $5\%$ of that in \cite{Yang2016a}, which reduces the scattering opacity by a factor of about 20. As a result, we expect the polarization here to be $\sim 27/20=1.35$ times that in Yang et al., which is close to our result. 
The polarization pattern starts to deviate more from that of the optically and geometrically thin case as the optical depth increases, as we show next.


The optical depth can be varied in several ways, including through the disk density, grain 
properties, and the observing wavelength. Here, we will focus on the effects of varying the 
optical depth through the density scale $\rho_0$, increasing it by a factor of 10 (Model B in 
Table~\ref{tab:model}) and $10^2$ (Model C), respectively.  The results are shown in Fig.3, 
which contrasts the total intensity distributions and polarization patterns for the three cases of
different densities. In Model B where the density is increased by a factor of 10 over that of 
Model A, the characteristic absorption optical depth remains well below unity ($\tau_{\rm c,abs}
=0.136$), although this is no longer true for the scattering optical depth ($\tau_{\rm c,sca}=
1.17$). The higher optical depth leads to a drastic asymmetry in the distribution of the polarized
intensity, with the near side (to the right of the major axis) much brighter than the far side (to
the left of the major axis; see Fig.~\ref{fig:image}e). This happens despite the fact that the 
total intensity $I$ is slightly higher on the far side than the near side (Fig.~\ref{fig:image}d),
because of the disk geometry, especially the finite angular thickness of the disk (see Fig.~
\ref{fig:loc_incl} and associated discussion below). It follows immediately that the light from 
the near side is much more polarized than that from the far side, as shown in Fig.~
\ref{fig:image}f, where the polarization vectors are plotted with the vector length proportional 
to the polarization degree. Note that most polarization vectors in the central region are no 
longer parallel to the minor axis as in the optically and geometrically thin case 
\citep{Yang2016a}: they
become significantly rotated with respect to the minor axis direction. 
The rotation is 
especially evident at locations along the major axis where the polarization orientation rotates 
counterclockwise from the minor axis in one direction in the top hemisphere and in the opposite (clockwise) direction in 
the bottom. This bifurcation in polarization orientation is a major consequence of the higher 
optical depth in a disk of significant angular thickness. It
becomes even more apparent in Model C, where the disk density is increased by another factor of 10
(see Fig.~\ref{fig:image}i), so that both the absorption and scattering optical depths at the 
characteristic radius exceed unity ($\tau_{\rm c, abs}=1.36$ and $\tau_{\rm c,sca}=11.7$). In this
densest disk that is optically thick to both absorption and scattering, the polarized intensity be
comes ``kidney'' shaped (Fig.~\ref{fig:image}h), showing extreme near-far side 
asymmetry . The asymmetry in total intensity also becomes more prominent (Fig.~\ref{fig:image}g).

\begin{figure}
	\centering
	\includegraphics[width=\columnwidth]{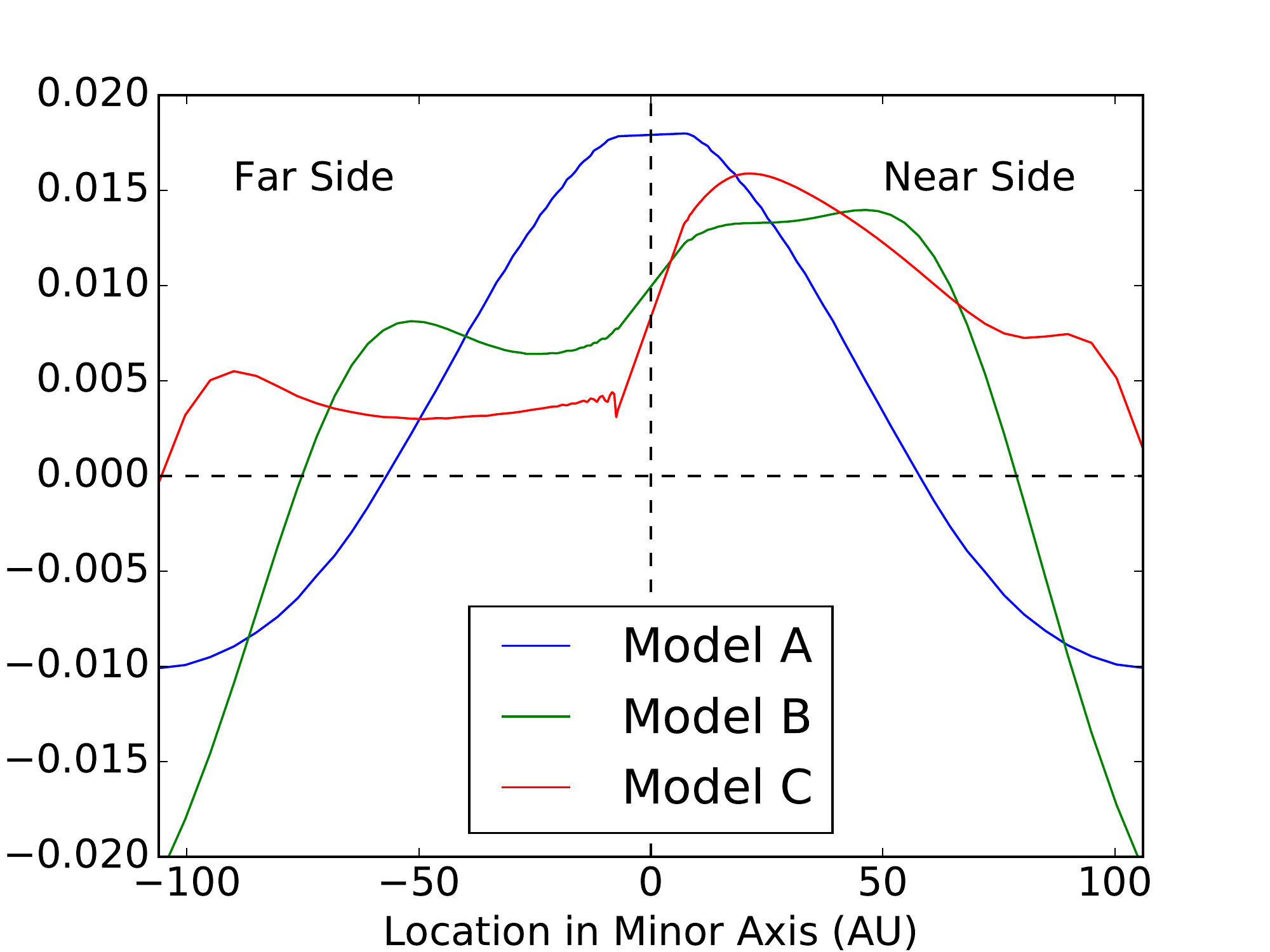}
	\includegraphics[width=\columnwidth]{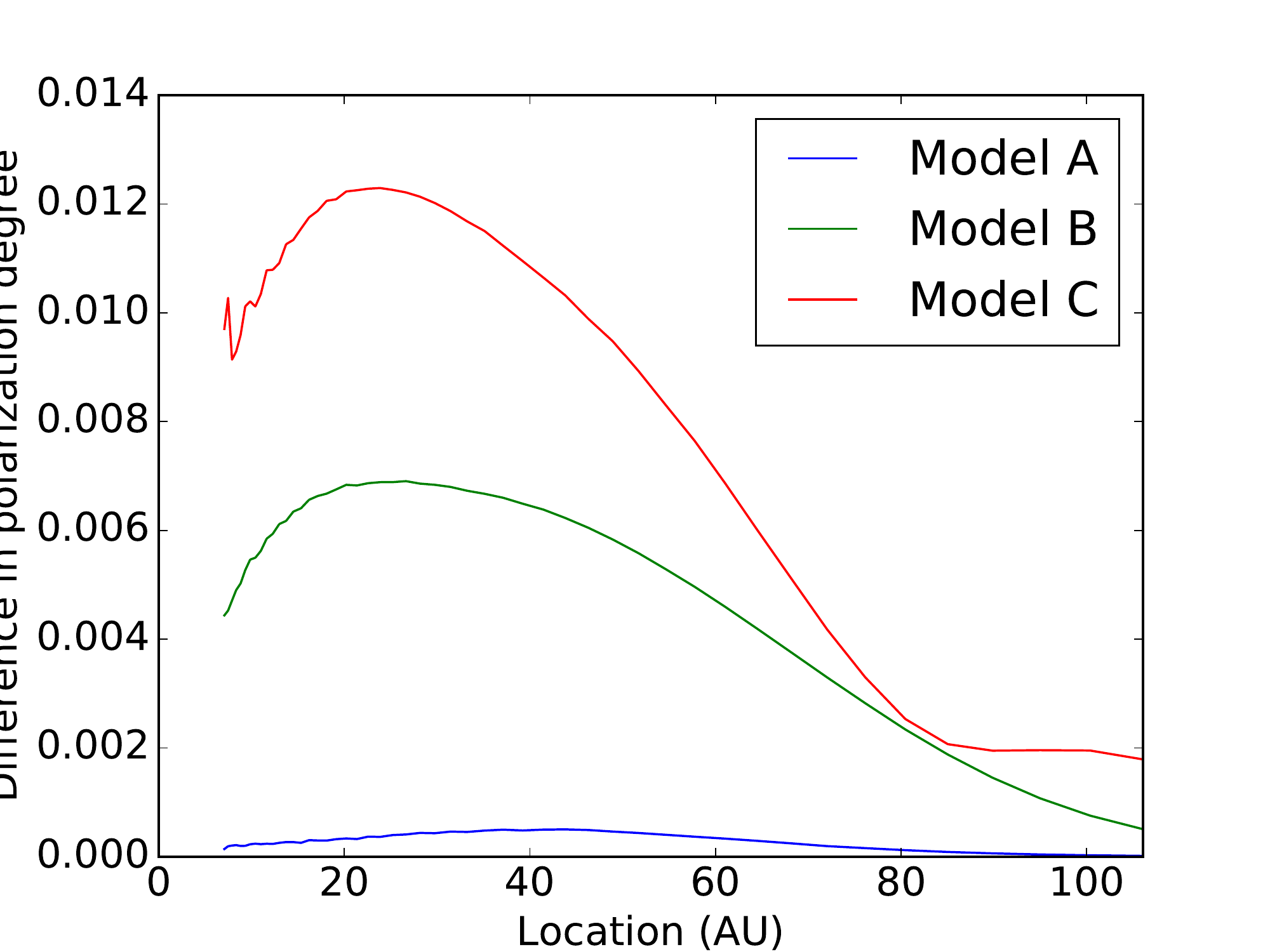}
	\caption{Top panel shows the polarization degree along the minor axis for the three
	models with different density scales, with the near side to the right of the
	central vertical line and the far side to the left. Positive polarization degree 
	indicate a polarization along minor axis, whereas a negative value corresponds to a 
	polarization along major axis. Bottom panel shows the difference
	in polarization degree for pairs of points along minor axis plotted against distance
	from the center of the disk for the same three models. Note the increasing near-far side asymmetry as the density (and thus optical depth) increases. }
	\label{fig:asym}
\end{figure}

\begin{figure}
	\centering
	\includegraphics[width=\columnwidth]{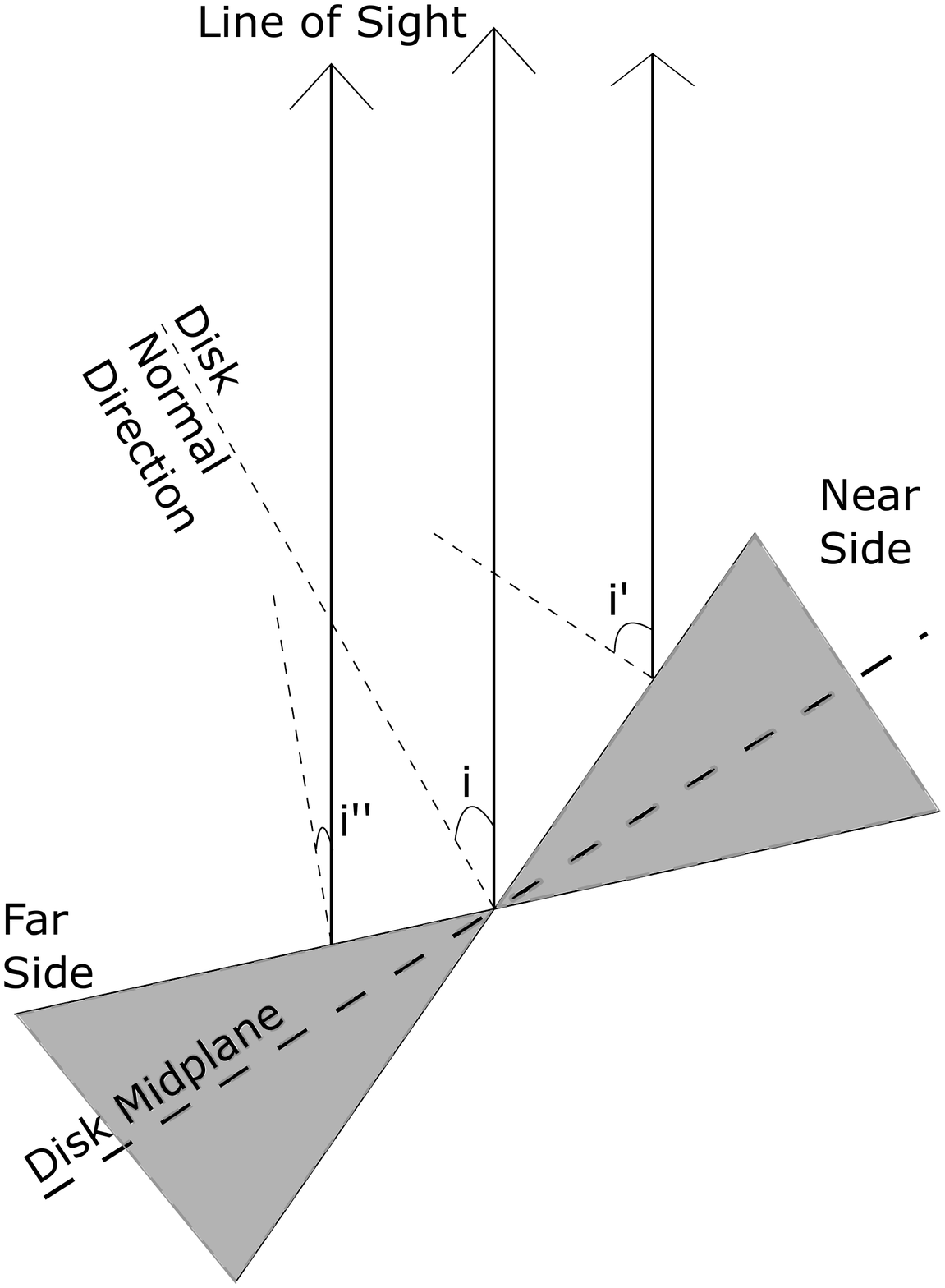}
	\caption{Illustration of the ``local inclination angle'' of the disk surface to the line of sight. The gray shaded region represents
	the disk. The line of sight makes a larger angle to the normal of the disk surface on the near side (denoted by $i'$) than that on the far side ($i''$). This difference in local inclination angle is the main reason for the near-far side asymmetry in the polarized intensity for optically thick disks.}
	\label{fig:loc_incl}
\end{figure}

The near-far side asymmetry is further quantified in Fig.~\ref{fig:asym}, where the
polarization degree along the minor axis is plotted against the distance from the center
for all three models in the top panel. A negative value for the polarization degree means that the polarization is along the major axis rather than the minor axis direction. It is clear that the polarization degree is very symmetric in the most optically thin Model A. As the optical depth increases, the polarization degree on the near side
of the disk (the part with positive distance from the center) stays high; indeed, it becomes larger than the optically thin case at some locations. The polarization degree on the far side (the part with
negative distance from the center), on the other hand, is significantly reduced, by a factor of $\sim 2.5$ for the moderately optically thick Model B and $\sim 4$ for the most optically thick Model C. This increase in asymmetry with optical depth is shown even more explicitly in the bottom panel of Fig.~\ref{fig:asym},
where we plot the difference between the polarization degrees at pairs of symmetric points
on the minor axis against the distance from the center. Note
that the difference is smaller at larger distances, especially beyond the characteristic
radius $R_c=79$~AU, where the density (and thus column density) drops precipitously. This
trend is consistent with the expectation that the asymmetry is controlled mainly by the
optical depth for a given (dust) disk geometry.

We can understand the role of the optical depth in breaking the near-far side symmetry using the semi-infinite
analytical model presented in \S~\ref{subsec:semiinfinite} in the following way. In the case of a highly optically thick disk,
the light observed along any line of sight comes from a thin layer near the disk surface, making the
emitting region effectively plane-parallel and semi-infinite locally. In this case, the degree of
polarization produced by scattering is sensitive to the inclination of the local disk surface to the line
of sight, as illustrated in Fig.~\ref{fig:pincl}. This local inclination angle is different for the near and far side
for any disk of a finite angular thickness, as illustrated in Fig.~\ref{fig:loc_incl}. Specifically, for a pair of points
located symmetrically on either side of the center along the minor axis, the inclination angle for the
point on the near side ($i^{\prime}$ in the cartoon) is larger than that on the far side ($i^{\prime\prime}$).
The larger inclination angle leads to a higher degree of polarization, unless the angle is close to
$90^\circ$. In the latter case, the situation is more complicated, because the light on the near side would come increasingly from the outer (radial) edge of the disk rather than its (top) surface.
Note that for the same symmetric pair of points, the one on the far side of the disk surface is
closer to the star (see Fig.~\ref{fig:loc_incl}), and is thus brighter in total intensity $I$ because of a higher
temperature. This naturally explains the near-far asymmetry in the total intensity,
which is most prominent in the most optically thick case and is in a sense opposite to that
of the asymmetry in the polarized intensity (see Fig.\ref{fig:image}). In other words, the near side of an optically thick disk has a higher polarized intensity despite the fact that its total intensity is lower.
These gradients in intensity and polarized intensity together results in much higher polarization 
fractions on the near side of the disk than the far side of the disk.

%
%


\begin{figure}
	\centering
	\includegraphics[width=\columnwidth]{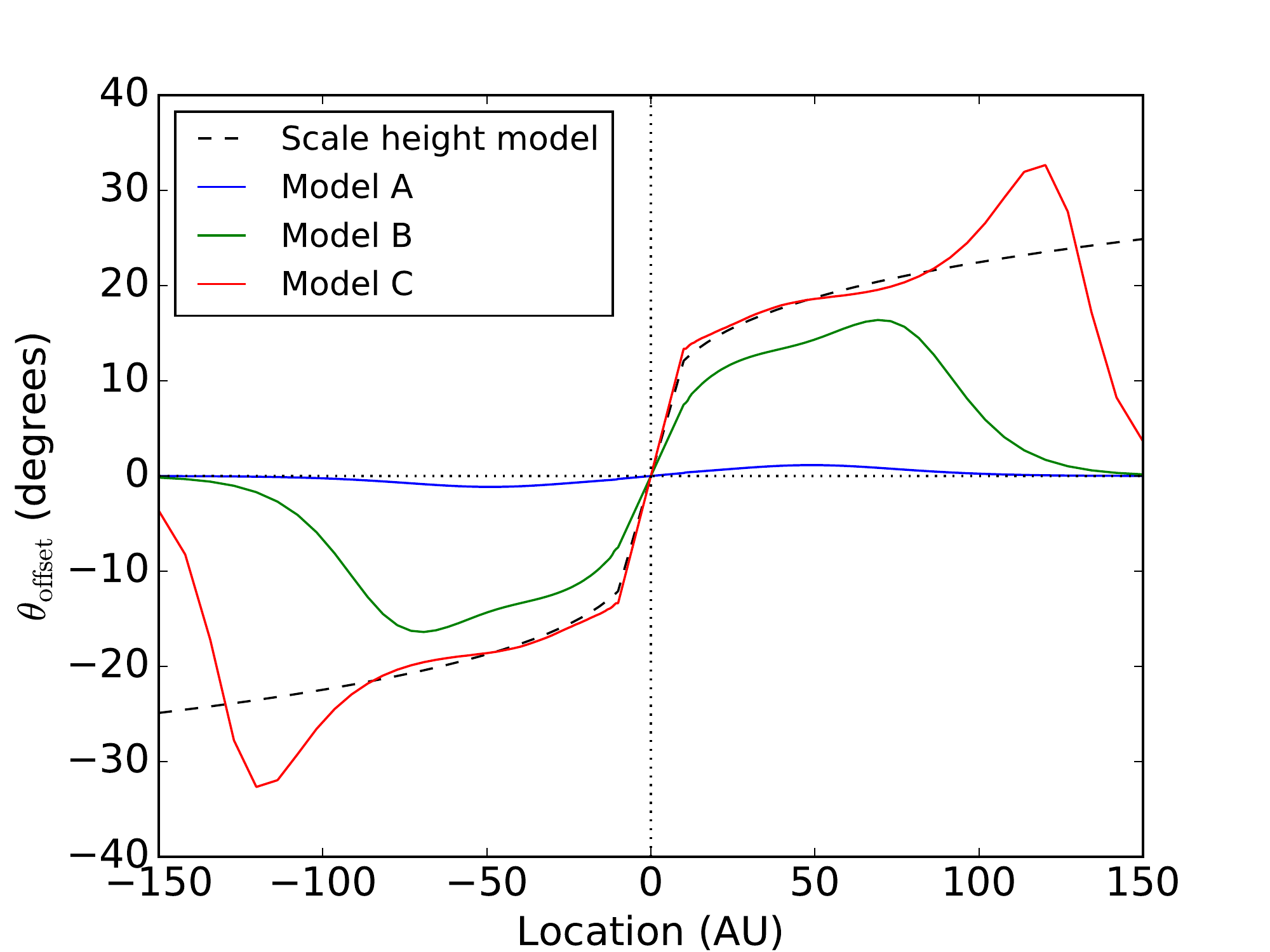}
	\caption{Bifurcation in polarization orientation along the major axis. Plotted are the offset angle $\theta_{\rm offset}$ between the polarization vectors and the direction of the minor axis as a function of distance from the center along the major axis for the three models with different density scales (Models A-C). The blue dashed line shows the expectation based on the shape of the disk surface at one scale height. }
	\label{fig:angles}
\end{figure}

Besides the near-far side asymmetry in polarized intensity, another important optical depth effect is the bifurcation in polarization orientation along the major axis. In the geometrically and optically thin case
discussed in \cite{Yang2016a}, the polarization
vectors at locations on the major axis are roughly uni-directional (parallel to the
minor axis; see also Fig.~\ref{fig:image}c). As the optical depth increases, the vectors start
to rotate more and more away from the minor axis direction and in opposite directions for the upper and lower hemispheres (see Fig.~\ref{fig:image}f and \ref{fig:image}i). We will
denote the angle between the polarization direction and the minor axis by $\theta_{\rm offset}$.
This effect is shown more quantitatively in Fig.~\ref{fig:angles}, where the offset angle $\theta_{\rm offset}$ is plotted against the distance along the major axis for the three models with different density scales. It is clear that $\theta_{\rm offset}$ is close to zero for the most optically thin case, but becomes more significant for denser disks. In addition, for each disk, 
$\theta_{\rm offset}$ decreases quickly at large radii, where the surface density also quickly decreases. Both of these trends support the notion that the offset is an optical depth effect. Furthermore, the offset angle is positive in the upper hemisphere (defined as rotating counterclockwise from the west in the right panels of Fig.~\ref{fig:image}) but negative in the lower hemisphere (rotating clockwise from the west), showing explicitly the bifurcation of polarization orientation along the major axis.
In other words, the position angle of the polarization vector is inclined in a mirror-symmetric manner 
with respect to the minor axis of the disk that is assumed to be intrinsically axisymmetric. 

This effect can also be understood with the help
of the analytical model developed in \S~\ref{subsec:semiinfinite}, where we showed that, for a semi-infinite
slab, the polarization direction is along the local ``minor axis'' (see Fig.~\ref{fig:cartoon}). As we discussed earlier, in
a highly optically thick disk, the light we see along any line of sight comes from a small
patch on the disk surface facing the observer that is effectively plane-parallel and
semi-infinite. The light from that patch is therefore expected to be polarized along the
direction of the {\it local} minor axis, which lies at the intersection of the plane
of the sky and the plane containing the normal of the local disk surface and the line of
sight. For a disk of finite angular thickness, this {\it local} minor axis is different
from the {\it global} minor axis, which is defined by the normal of the disk {\it mid-plane} (or the rotation axis of the disk). This is especially true for a flared disk, where the normal of the local disk surface becomes increasingly misaligned with respect to the global (rotation) axis as the radius increases. The increasing misalignment between the local disk normal and the global axis leads to an increasingly large offset angle $\theta_{\rm offset}$ between the local minor axis and the global minor axis in the plane of the sky, especially at locations along the global major axis. This is illustrated in Fig.~\ref{fig:angles}, where the variation of the offset angle $\theta_{\rm offset}$ is plotted as a function of radius (the dashed blue line) as one moves away from the center along the global major axis on a surface one scale height $H(R)$ above the disk mid-plane. The curve agrees remarkably well with that of the most optically thick case inside the characteristic radius $R_c=79\rm\,AU$, which supports our geometric interpretation of the offset. Beyond $R_c$, the column density drops quickly, making the disk increasingly optically thin and the surface of one scale height increasingly less representative of the $\tau=1$ surface. The disagreement beyond $R_c$ is therefore to be expected; it strengthens (rather than weakens) our interpretation.

Note that, in Fig.~\ref{fig:angles}, we chose to plot the offset angle
$\theta_{\rm offset}$ only along the major axis, even though the effect is not
limited to these locations. There are two reasons for this choice. First, the
polarization orientations there are all along the minor axis (i.e.,
$\theta_{\rm offset}=0$) in optically thin limit, which makes it easier to highlight
the optical depth effects. Just as importantly, it is easier to check the results
with simple geometric disk surface models on the major axis than elsewhere, as we
have done in Fig.~\ref{fig:angles} (the dashed blue line). We reiterate that the polarization orientations on the major axis have mirror symmetry with respect to the minor axis (assuming an
intrinsically axisymmetric disk), with the polarization line segments rotating away
from the direction of the minor axis by an acute angle counterclockwise in one
hemisphere and clockwise in the other (see Fig.~\ref{fig:image}f and i). If such a unique pattern of bifurcation in polarization orientation is observed, we can potentially infer the dependence of the dust scale height on the radius, provided that the disk is optically thick enough.

\subsection{Monte Carlo radiative transfer and the effects of multiple scattering}
\label{subsec:mc}

So far, we have determined the disk polarization by solving the radiative transfer equation for polarized light using the formal solution under the single scattering approximation. Strictly speaking, the approximation is only valid when the disk is optically thin to scattering. Otherwise, multiple scattering may be important. In this subsection, we gauge the potential effects of multiple scattering on the disk polarization, especially the near-far side asymmetry, through Monte Carlo simulation using the publicly available code \texttt{RADMC-3D}\footnote{\texttt{RADMC-3D} is available at
\url{http://www.ita.uni-heidelberg.de/~dullemond/software/radmc-3d/}}.
For the three models presented in this subsection, we allow photons to be 
polarized between scattering with the scattering phase matrix calculated with 
Mie theory \citep{BH83}. Each model uses 1.28 billions photon packages. 

\begin{figure*}
    \centering
    \includegraphics[width=0.8\textwidth]{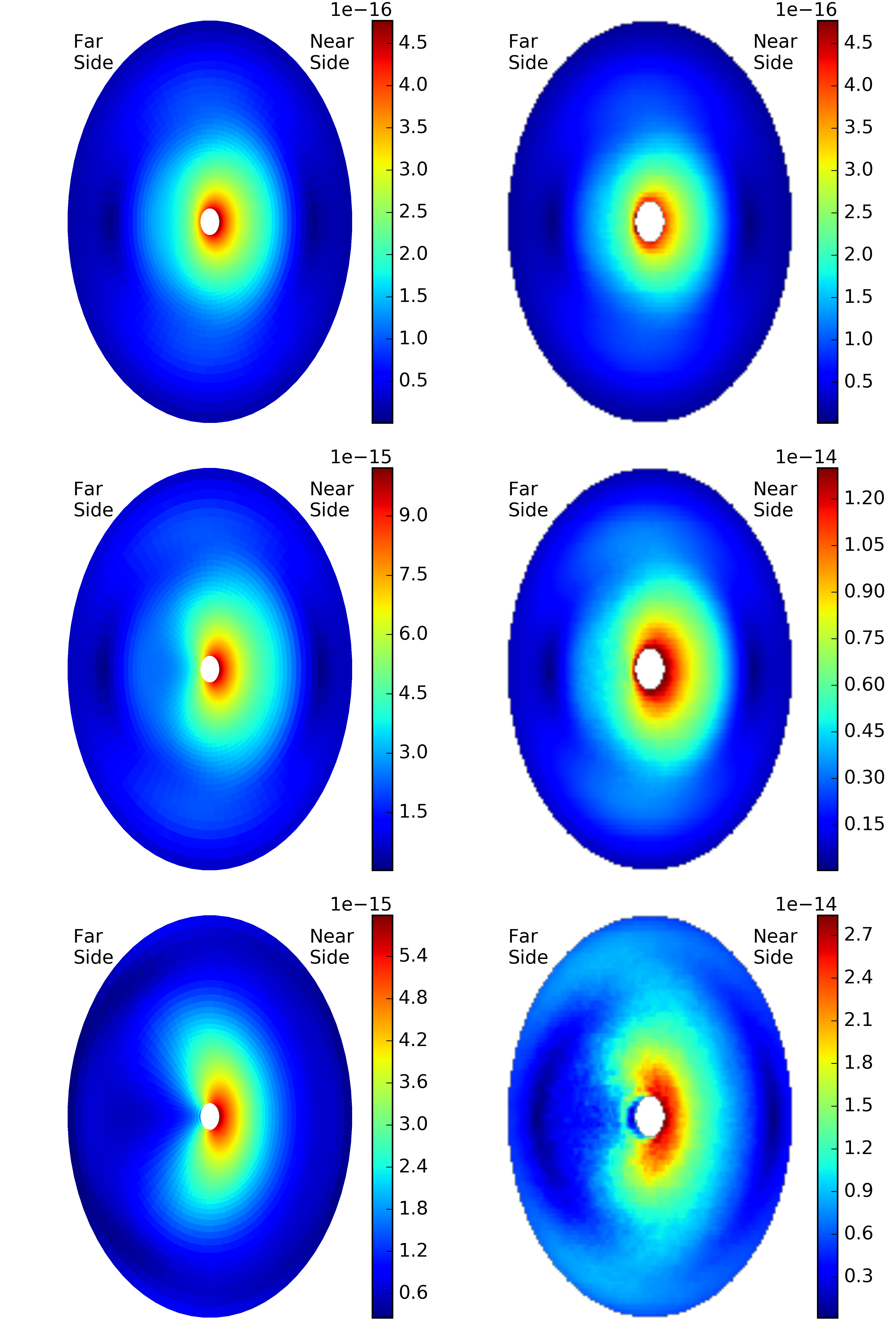}
    \caption{Comparison of the results from formal solution under single scattering
    approximation (left panels) and Monte Carlo simulations including multiple
    scattering (right panels). Plotted are the polarized intensity for a representative
    disk inclined by an angle $i=45^\circ$ to the line of sight, with three different
    grain sizes: 10~${\rm\,\mu m}$ (Model E; top panels), 37.5~${\rm\,\mu}$ (Model D; middle) and
    100~${\rm\,\mu m}$ (Model C; bottom). From top to bottom, the scattering optical depth at the characteristic radius $R_c$ are
    $1.04\times 10^{-2}$, $0.56$, and $11.7$, respectively.
    Intensities are in unit of
    $\rm erg\cdot s^{-1}\cdot sr^{-1}\cdot cm^{-2}\cdot Hz^{-1}$.
    The color scales were adjusted to best show the morphology of the polarized intensity.}
    \label{fig:diffa}
\end{figure*}

To facilitate comparison with the calculations discussed above, we will adopt the same
physical model for the disk as before, focusing in particular on the highest density
case of $\rho_0=1.13\times 10^{-14}$~g~cm$^{-3}$, where the potential effects of
multiple scattering are expected to be the largest. We will consider a set of three   
models with three different (spherical) grain sizes, $10\rm\, \mu m$ (Model E in Table~\ref{tab:model}, $37.5\rm\, \mu m$ (Model D),
and also the $100 \rm\,\mu m$ (Model C) as before. The grain sizes are chosen so that the scattering
opacities span a large range in value (3 orders of magnitude), although the  absorption
opacities remain rather constant (they differ by $\sim 40\%$ or less). In other words, these
models have similar absorption optical depths but vastly different scattering optical depths,
with values at the characteristic radius $R_c$ of $\tau_{c,sca}=1.04\times 10^{-2}$, $0.56$,
and $11.7$, respectively.
The smallest grain size is chosen
to provide an independent check on the results of the Monte Carlo simulations, which
should be close to those obtained through the formal solution under the single
scattering approximation since the scattering optical depth is much smaller than unity
throughout the disk. In the top panels of Fig.~\ref{fig:diffa}, we plot the distributions
of the polarized intensity of the 10~$\rm\,\mu m$ grain case (Model E) computed from these two methods side by side. We can see that
they do look very similar qualitatively (and quantitatively, see below), which adds confidence to the results from both methods.
Interestingly, there is a significant near-far side asymmetry even in this case of very small
scattering optical depth. The implication is that a high {\it absorption} optical depth is
sufficient by itself to generate a strong near-far side asymmetry. This is not
surprising since a large absorption optical depth is enough to limit the source region
of the polarized (scattered) light to the surface of the disk facing the observer, which
has a near-far side asymmetry as viewed by the observer (see illustration in Fig.~\ref{fig:loc_incl}).
However, in this particular case, the degree of polarization in the region of most
interest (within the disk characteristic radius $R_c$) is of order $2\times 10^{-4}$,
which is well below the detection limit of even ALMA.

To obtain a higher degree of polarization, we first increase the grain size to
37.5$\rm\,\mu m$ (Model D), which corresponds to an increase of the scattering optical depth by a
factor of $\sim 50$. An increase of the polarization degree by a similar factor is
expected (to about $\sim 1.0\%$). The actual polarization degree is somewhat less,
reaching a value about half of the expected one. The discrepancy
is most likely due to the fact that the disk is starting to become optically thick to
scattering at the characteristic radius (with $\tau_{\rm c,sca}=0.56$) and
more so at small radii.
%
%
As the grain size increases further to 100$\rm\,\mu m$ (Model C), the scattering opacity depth
increases by another factor of $\sim 20$, to $\tau_{\rm c,sca}=11.7$.
Even in this most extreme case where the disk is optically thick to both absorption and
scattering, the spatial structure of the polarization patterns from the two methods (the formal solution and RADMC-3D) remain qualitatively similar, as shown in the bottom panels in Fig.~\ref{fig:diffa}. In particular, the inclusion of
multiple scattering in the RADMC-3D case does not erase the near-far asymmetry in the
distribution of the polarized intensity; the ``kidney'' shaped distribution broadly
resembles that obtained with formal solution without multiple scattering.

\begin{figure*}
\includegraphics[width=\textwidth]{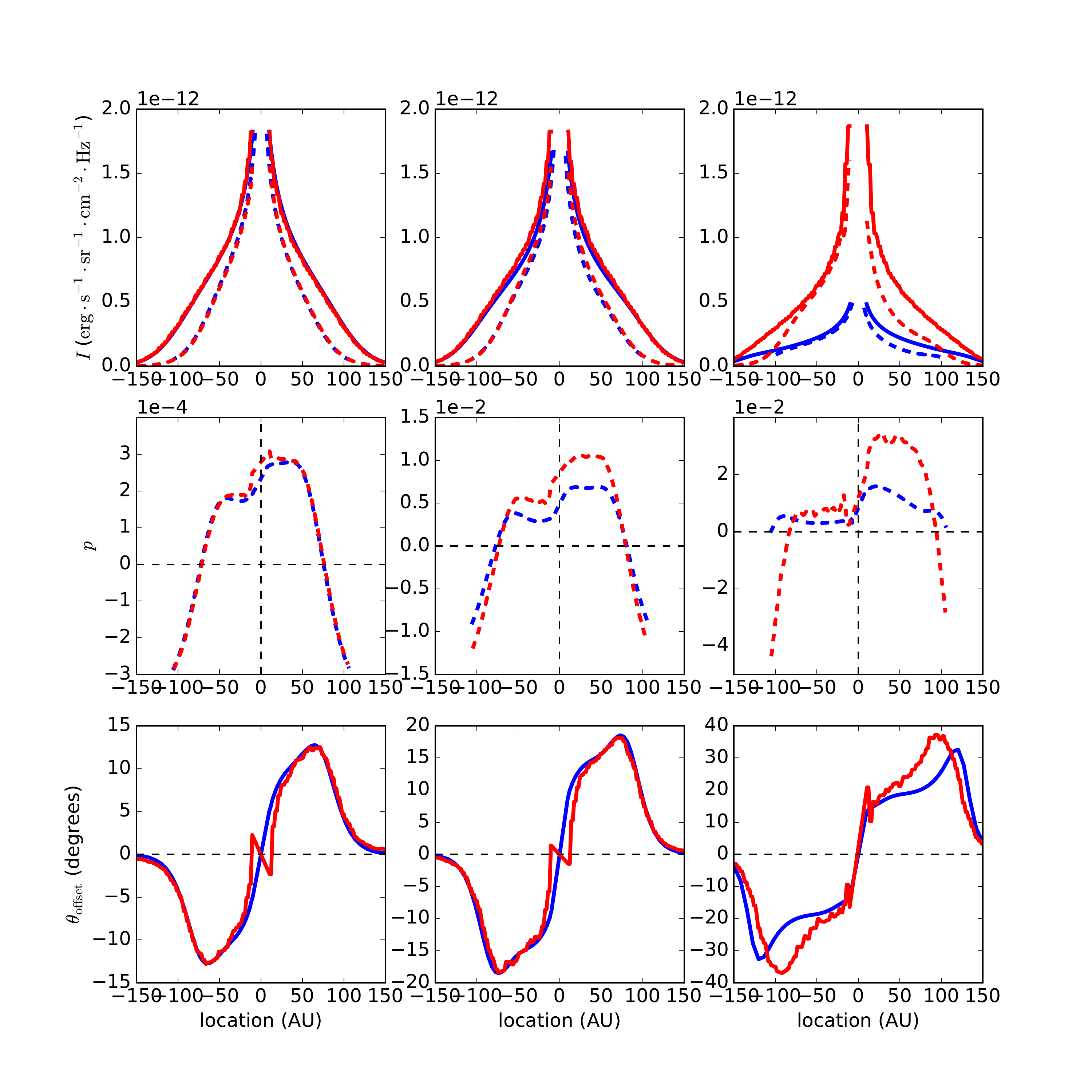}
\caption{Quantitative comparison of the results from formal solution under single
scattering approximation (blue curves) and Monte Carlo simulations including multiple
scattering (red curves). Plotted are the total intensity $I$ along both the major and minor
axis (top panels), the polarization degree $p$ along the minor axis (middle), and the offset
angle $\theta_{\rm offset}$ between the polarization orientation and the minor axis along the
major axis (bottom), for three grain sizes: $10\rm\,\mu m$ (Model E; left panels), $37.5\rm\,\mu m$
(Model D; middle), and $100\rm\, \mu m$ (Model C; right). Results on major axis and minor axis are plotted in solid lines and dashed lines, respectively.}
\label{fig:comp}
\end{figure*}

To compare the results from the two methods more quantitatively, we plot in
Fig.~\ref{fig:comp} the distributions of the total intensity $I$ along both the
major and minor axis (top panels), the polarization degree $p$ along the minor axis
(middle), and the offset angle $\theta_{\rm offset}$ between the polarization
orientation and minor axis along the major axis (bottom).
The three columns show models with grain sizes of
$10\rm\,\mu m$ (Model E), $37.5\rm\,\mu m$ (Model D) and $100\rm\, \mu m$ (Model C), respectively. Red curves are
from RADMC-3D and blue curves from the formal solution. It is clear that there is good agreement in the case of the
smallest grain (the left panels; optically thin to scattering) for all
three plotted quantities ($I$, $p$ and $\theta_{\rm offset}$). Multiple scattering
does have some effects on the polarization degree $p$ for the case of intermediate grain size (middle column), increasing it by a factor of $\sim 30-50\%$ (see the middle panel of the figure). As expected, it affects the case of the largest grain (and the highest scattering optical depth) the most. In particular, multiple scattering increases the total intensity along both the major and minor axes by up to a factor of $\sim 4$ compared to that obtained under the assumption of single scattering. This is because photons are heavily extincted due to high scattering optical depth, and the bulk of such extincted photons would reappear through the disk surface and be observed as (more polarized) scattered photons when multiple scattering is taken into account but not in the single scattering limit. The larger number of (polarized) scattered photons toward the observer naturally explains why the polarization degree is significantly higher with multiple scattering than without, especially on the near side of the inclined disk (see the middle-right panel). In any case, multiple scattering does not erase the near-far asymmetry; if anything, the asymmetry is enhanced by multiple scattering, especially in the case of larger grains. Moreover, multiple scattering events have little effect on the optical depth effect of bifurcation in polarization orientation along the major axis, as measured by the offset angle $\theta_{\rm offset}$, as shown in the bottom panels of Fig.~\ref{fig:comp}. This is not too surprising because the orientation of the polarization is expected to be
determined mainly by the projected normal direction of the local disk surface in the plane of the sky in optically thick regime (see \S~\ref{subsubsec:results} and Fig.~\ref{fig:angles}), which is relatively unaffected by multiple scattering.

\section{Disk polarization as a probe of dust settling}
\label{sec:settling}

One of the most striking features of the scattering-induced polarization in an inclined, optically and geometrically thick dust disk is the near-far side asymmetry in polarized intensity. It comes about because (1) the observed photons come from the surface layers of the optically thick disk, and (2) the surface on the near-side is viewed more edge-on than that on the far-side because of the finite angular thickness of the (dust) disk (see Fig.~\ref{fig:loc_incl}). Here, we first demonstrate explicitly that the near-far side asymmetry disappears in a geometrically thin dust disk (\S~\ref{subsec:thin}), and then discuss the potential of using the near-far side asymmetry as a probe of the thickness of the dust layer responsible for the scattering (i.e., dust settling; \S~\ref{subsec:evolution}) and as a way to differentiate dust scattering from other mechanisms for producing disk polarization in (sub)millimeter (such as direct emission from magnetically or radiatively aligned grains, \S~\ref{sec:mechanism}).

\subsection{Dependence of near-far side asymmetry on dust settling}
\label{subsec:thin}

\begin{figure*}
    \centering
    \includegraphics[width=\textwidth]{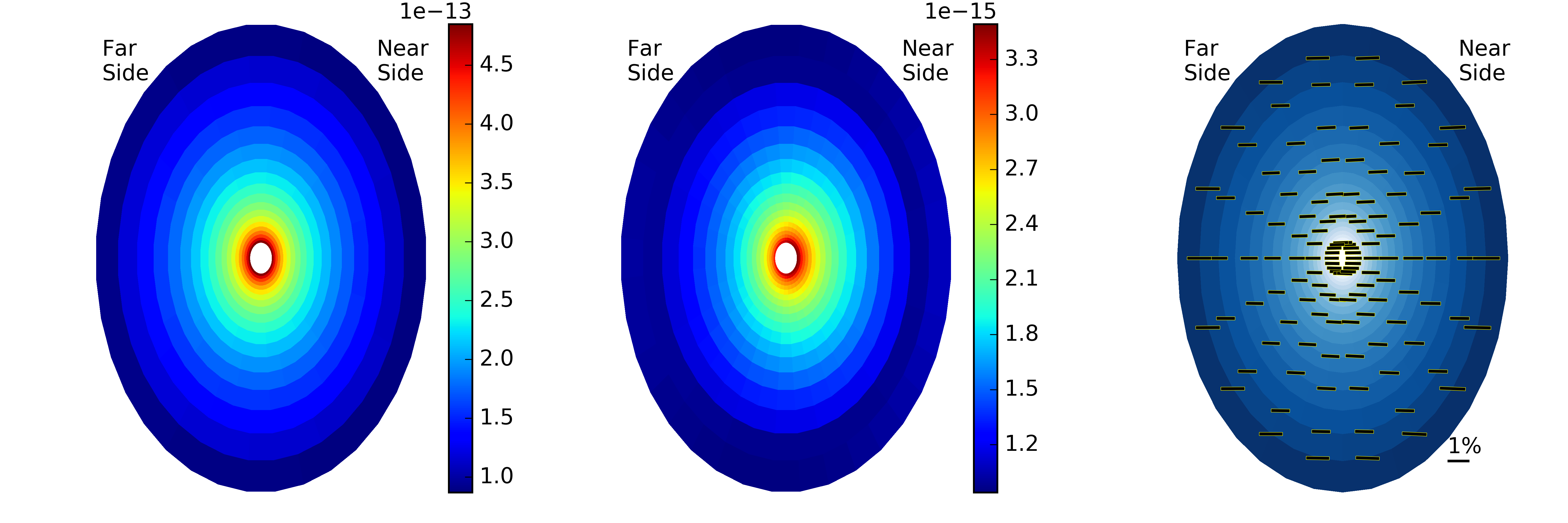}
    \caption{Total intensity (left), polarized intensity (middle), and
    polarization vectors (right) for geometrically thin but optically thick
    disk model (Model F). Intensities are in unit of
    $\rm erg\cdot s^{-1}\cdot sr^{-1}\cdot cm^{-2}\cdot Hz^{-1}$. }
    \label{fig:thinner}
\end{figure*}

To illustrate the dependence of the near-far side asymmetry on the thickness of the layer of (large) grains responsible for the scattering, we repeat the most optically thick model discussed in \S~\ref{subsec:fs} (Model C in Table~2) but with the (dust) scale height reduced by a factor of 10. The (dust) density is increased by a factor of 10 correspondingly, to keep the column density and thus the optical depth the same (Model F). The results are shown in Fig.~\ref{fig:thinner}, where the total intensity, polarized intensity, and polarization vectors are plotted. They can be directly compared with the results shown in
Fig.~\ref{fig:image}g, Fig.~\ref{fig:image}h, and Fig.~\ref{fig:image}i for the corresponding thicker (dust) disk case (Model C). Several features are immediately apparent from the comparison. First, the two cases have similar scales for the total intensity, which is expected since they have the same distributions of column density and temperature. A major difference is that the disk is nearly symmetric in total intensity and polarized intensity in the geometrically thinner case (Model F) but significantly brighter in total intensity on the far-side and in polarized intensity on the near-side in the geometrically thicker case (Model C). This is not surprising because, as the (angular) thickness of the dust disk shrinks, the surfaces of the near and far sides of the disk become more symmetric with respect to the line of sight (see Fig.~\ref{fig:loc_incl}). Furthermore, both surfaces are closer to the disk midplane, reducing the difference between the global minor axis (defined by the disk axis or midplane) and the local minor axis on the disk surface facing the observer, which makes polarization vectors more aligned with the (global) minor axis, as shown in the right panel of Fig.~\ref{fig:thinner} (see Fig.~\ref{fig:image}h for comparison). In other words, the degree of bifurcation in polarization orientation along the major axis is also reduced for a geometrically thin disk. These differences can in principle provide a way to determine whether the large grains responsible for the scattering-induced polarization are settled to the disk midplane or not.

\subsection{Evidence for evolution of dust settling over time}
\label{subsec:evolution}

Dust settling is an important process to study because it tends to increase the local concentration of the dust-to-gas ratio near the midplane and may play a crucial role in the formation of planetesimals and ultimately planets.
Physically, whether grains would settle toward the disk midplane or not depends on their sizes, the level of turbulence and the distribution of gas density in the disk, as well as the age of the system. It is expected to occur preferentially for large grains in relatively quiescent, evolved disks. We will first review anecdotal evidence from (unpolarized) dust continuum imaging of edge-on disks that large grains are not yet settled in Class 0 disks, but become more so at later (Class I and II) stages (\S~\ref{subsec:edge-on}). This is followed by complementary evidence for a similar trend from the currently available disk polarization observations of inclined disks (\S~\ref{subsec:asymmetry}).

\subsubsection{Evolution of dust settling in edge-on disks from dust imaging}
\label{subsec:edge-on}

High resolution observations of (unpolarized) dust continuum, especially with ALMA, have shown that the grains in the two best-studied edge-on Class 0 disks, HH212 and L1527, are vertically distributed. In the case of HH212, the evidence comes from long-baseline ALMA continuum observation in Band 7 (0.85~mm) that resolved the disk vertical structure with a 0.02$^{\prime\prime}$ resolution \citep{Lee2017}. It reveals a dark lane near the disk midplane sandwiched between two brighter regions. Preliminary modeling of this ``hamburger-shaped" continuum indicates that the dust layer responsible for the detected emission at 0.85~$\mu$m has a scale height of $\sim 10$~AU at a radius of $\sim 40$~AU, making the dust disk rather thick geometrically. The dust opacity index $\beta$ was estimated at $0.6$ between $0.85$ and $1.4$ mm wavelengths, which, taken at the face value, would indicate significant grain growth, possibly to millimeter sizes or larger  \citep{Testi2014}. These observations appear to indicate that the grains in the HH212 Class 0 disk have already grown significantly but not yet settled toward the mid-plane. The same picture appears to hold for the edge-on Class 0 disk L1527 as well, which has an estimated (total) thickness (full width at half maximum) of $\sim 30$AU at a radius of 50~AU based on ALMA Band 7 continuum observations at 0.8~mm (\citealt{Sakai2017}; see also {Aso et al. 2017 submitted}). These two examples provide a strong motivation to study the scattering-induced polarization from dust grains that are yet to settle to the mid-plane, as we have done in \S~\ref{sec:numerical}.

The situation appears different for more evolved disks. For example, for the iconic disk of the Class I/II object HL Tau, \cite{Pinte2016} was able to infer that the large grains responsible for the (sub)millimeter continuum emission are settled to the mid-plane, based on the lack of azimuthal variation of the width of the gaps in the significantly inclined disk \citep{ALMA2015}. Other possible examples include the disk of the well-studied Class II object HH30, which remains unresolved in the vertical direction by ALMA observations at 1.3~mm ({F. Menard et al. 2017, submitted}), and the edge-on disk of the so-called ``Flying Saucer," where large grains appear to be settled to the midplane \citep{Guilloteau2016}. Although the number of edge-on disks with high resolution ALMA data remains small, the available evidence is suggestive of a picture where grains responsible for (sub)millimeter continuum emission remain vertically distributed in the youngest (Class 0) disks, but become more settled at later times.

The tentative trend for the evolution of dust settling is perhaps to be expected, for two reasons. First, it takes time for the grains to grow and settle, which tends to favor dust settling in older sources. Perhaps more importantly, the youngest disks are inferred to accrete at much higher rates than older disks (by one to two orders of magnitude or more, e.g., \citealt{Yen2017}). If the accretion is driven by turbulence (induced by magneto-rotational instability or some other means), the level of turbulence in the youngest disks must be much higher, which could plausibly stir up the grains enough to prevent them from rapid settling toward the mid-plane (e.g., \citealt{Fromang2009}). If this is indeed true, it would have far-reaching implications for the timing of planet formation: if planetesimals are not formed in earnest until grains have settled, it would be difficult to form planets from planetesimals during the Class 0 phase. A corollary would be that there should not be any planet-induced rings and gaps on Class 0 disks\footnote{Rings and gaps would be smoothed out quickly by a high level of turbulence in Class 0 disks, potentially making this supposition difficult to test.}. Given its significance, it is important to probe dust settling using an independent method. We believe that disk polarization has the potential of being developed into one such method.
%
%
%
%
%

\subsubsection{Evolution of dust settling from polarization of inclined disks}
\label{subsec:asymmetry}

%
%

Although the field of disk polarization is poised for rapid growth in the ALMA era, the number of disks with resolved polarization detection remains small. Nevertheless, there is already some indication for the near-far side asymmetry in polarized intensity expected in an optically and geometrically thick (dust) disk. The best example to date is the inclined, optically thick disk of the famous massive protostar HH80-81 ({J. M. Girart, in prep}). ALMA observations revealed a well resolved polarization pattern that is roughly parallel to the minor axis close to the center and more azimuthal further out, broadly consistent with the pattern from scattering (\citealt{Yang2016a,Yang2016b,Kataoka2016a}; see Fig.~\ref{fig:image}). Interestingly, the near side of the disk, as inferred from its projection in the plane of the sky on the redshifted jet, is much brighter than the far side in polarized intensity, as expected from scattering-induced polarization with the (large) grains responsible for the scattering not yet settled to the mid-plane.

A similar asymmetry was also observed with ALMA in one of the youngest intermediate mass protostars, OMC3 MM6, with the polarized intensity much brighter on the near side (projected against the redshifted outflow) than the far side ({S. Takahashi et al. 2017, submitted}). In addition, high-resolution VLA observations revealed that the well-studied low-mass Class 0 protostar NGC 1333 IRAS 4A is significantly polarized at 8~mm, again with the near side brighter than the far side in polarized intensity \citep{Cox2015,Liu2016}. The source is so bright that it is believed to be optically thick out to 7~mm \citep{Liu2016}, which is consistent with the interpretation that the polarization comes from scattering, at least near the center (see also \citealt{Yang2016b}). Although it is conceivable that in each of the sources the asymmetry could be produced by an intrinsic feature on the disk, such a dust trap, it is statistically unlikely for such a trap to occur on the near side for all three cases. We conclude that the relatively sparse high-resolution polarization data currently available is consistent with the picture painted by the edge-on disks above, that, during the earliest phase of star formation, grains have already grown significantly to enable detectable polarization from scattering but have yet to settle toward the mid-plane, for both high-mass and low-mass protostars.

For more evolved objects, the situation can be quite different. For example, there is no clear indication of a near-far side asymmetry in the polarized intensity of the HL Tau (Class I/II) disk from the CARMA observations at 1.3~mm \citep{Stephens2014}, although the spatial resolution is rather limited. The lack of asymmetry is to be expected if the large grains are indeed settled to the mid-plane, as inferred from the shape of the gaps on the disk \citep{Pinte2016}. If the grains in the HL Tau disk have indeed settled, a prediction would be that there would not be any significant near-far side asymmetry even in ALMA Band 7 (the currently shortest wavelength available for polarization observations) where the disk is known to be optically thick. This prediction can readily be tested with ALMA observations. ALMA polarization data is available for the transition disk HD 142527 \citep{Kataoka2016a}, although its interpretation is complicated by the intrinsic asymmetry in the dust distribution. In any case, the observed pattern does not contradict in any obvious way the expectation that the large millimeter-emitting grains have settled in this evolved disk, although more detailed modeling is required to draw a firmer conclusion.

We therefore have two complementary methods of probing dust settling: vertically resolved (non-polarized) continuum imaging of edge-on disks and scattering-produced polarization in inclined disks. The first method is direct and can be interpreted easily with few assumptions but requires nearly edge-on systems (which are rare) and very high spatial resolution. The second method is less demanding in disk inclination (and hence is potentially applicable to more sources) and spatial resolution, but the polarized intensity tends to be much weaker than the non-polarized intensity. It does have the advantage of providing a tighter constraint on the grain size than direct imaging, since the scattering coefficient is much more sensitive to the grain size than the emission coefficient.



\section{Distinguishing mechanisms of disk polarization}
\label{sec:mechanism}

\subsection{Scattering vs emission by aligned grains}
\label{subsec:contrast}

Besides scattering, it is well-known that aligned (non-spherical) grains can also produce polarization through direct emission. Whether large grains responsible for (sub)millimeter emission can be aligned or not inside young star disks remains debated. \cite{Hoang2017} pointed out that, if there are enough iron atoms per cluster inside a grain (i.e., if the grain material is superparamagnetic enough), magnetic fields may align grains with sizes up to a few millimeters. In the case that the number of iron atoms is too small for magnetic alignment, it may still be possible for the grain to be aligned with respect to the direction of anisotropy in the radiation field through radiative torques \citep{Tazaki2017}. The contrast between the polarization patterns from scattering and magnetically aligned grains was already discussed in some depth in \cite{Yang2016b}. In this subsection, we will focus on the similarities and differences between those from scattering and  radiatively aligned grains.

Both polarization mechanisms depend strongly on the degree of anisotropy
in the radiation field: scattering of isotropic radiation does not produce
any polarization at all, and radiative torques are far weaker for isotropic
radiation field than for anisotropic ones \citep{Lazarian2007}. Radiatively
aligned (non-spherical) grains precess around the direction of the radiation
flux, with their long axis perpendicular to the flux direction \citep{Tazaki2017}. In the
simplest case of a face-on, axisymmetric disk, the radiative flux would be
in the radial direction, which would force the grains to align their long
axes along the azimuthal direction, producing an azimuthal pattern that is
similar to the pattern produced by scattering in a face-on disk (see the
upper-left panel of Fig.~2 of \citealt{Yang2016a}). However, there is
substantial difference between the two mechanisms for significantly
inclined (axisymmetric) disks: the radiatively aligned grains would produce
a radial polarization pattern, which is similar to the pattern produced by
scattering at large distances from the center but not near the center. For scattering, 
polarization vectors toward the center are more or less parallel to the minor axis,
especially for locations on the minor axis. This is an inclination effect
caused by the dependence of polarization degree on the scattering angle and
thus unique to the scattering-induced polarization (see
Fig.~\ref{fig:image}C and the lower-left panel of Fig.~2 of
\citealt{Yang2016a}). High resolution may be required to resolve the
central region to find this tell-tale sign for scattering.
%
%
%
%
%

It may be possible to distinguish polarization from scattering and radiatively aligned 
grains even in a face-on disk if the dust distribution is non-axisymmetric (i.e., dust traps). 
This is because the photons involved in producing the polarization at a given observed wavelength 
(e.g., (sub)millimeter) are quite different for the two mechanisms. For the  scattering-induced polarization, it is the anisotropy of the radiation field at the observed wavelength before the scattering that is responsible for the polarization at the same wavelength after the scattering. This is completely different from the case of radiatively aligned grains where the grain alignment is dominated by the anisotropy of the photons at wavelengths near the peak of the spectral energy distribution (SED) at the grain location, which are usually quite different from the observed wavelength. In particular, the radiation energy density in the bulk of a disk may be dominated by the radiation from the warmer surface layers, which typically have wavelengths much shorter than the mm/sub-mm probed by ALMA. If there is a difference in anisotropy between photons at the observation wavelength and near the SED peak, the polarization patterns produced by the two mechanisms would be different. For illustration, consider the transition disk around Oph-IRS 48. The $440\rm\, \mu m$ image observed with ALMA in Band 9 shows a ``kidney-shaped" concentration of large grains (i.e., a dust trap). The $18.7\rm\,\mu m$ emission detected by the Very Large Telescope is, on the other hand, much more uniformly distributed in the azimuthal direction \citep{vanderMarel2013}. If the radiation field near the SED peak has a distribution closer to that in the mid-infrared than in sub-mm, we would expect the (non-spherical) grains inside the ``kidney-shaped" dust trap to be aligned more or less azimuthally, with a corresponding azimuthal polarization pattern from direct thermal dust emission at mm/sub-mm wavelengths. Self-scattering of the mm/submm photons would produce a very different polarization pattern, with polarization vectors switching directions going from inside the dust trap to outside \citep{Kataoka2016b}. High resolution multi-wavelength (mid- and far-IR continuum and (sub)millimeter polarization) observations of this type of sources can help distinguishing these two mechanisms.

We note that aligned grains can produce polarization through both direct emission and scattering. In this paper, we have limited our treatment to scattering by spherical grains, with the emphasis on the optical depth effects. The combined polarization from both direct emission and scattering by aligned (non-spherical) grains has been considered in Yang et al. (2016b), but only in the optically and geometrically thin limit. These simplifications will be relaxed in a future investigation.

\subsection{Near-Far Side Asymmetry as Signpost for Scattering-Induced Polarization}
\label{subsec:signpost}

The presence (or absence) and sense of near-far side asymmetry in polarized intensity in inclined, optically and geometrically thick (dust) disks may be the key to distinguish the disk polarization induced by scattering from those by magnetically or radiatively aligned grains. In the simplest case of (rapidly spinning, effectively) oblate grains aligned with their shortest axes along a purely toroidal magnetic field, large near-far side asymmetry along the minor axis is not expected because the grains there are always viewed edge-on (and thus emit maximally polarized light) independent of the disk inclination angle. In the simplest case of oblate grains radiatively aligned with their shortest axes along a purely (spherically) radial direction, the observer would see the oblate grains located on the minor axis more face-on on the disk surface on the near side than that on the far side (see Fig.~\ref{fig:loc_incl} where the dashed lines on the near and far sides mark the mid-plane of the oblate grains). This would make the intensity of the polarized light emitted by the radiatively aligned grains weaker on the near-side than on the far-side, exactly the opposite of the scattering case. As discussed in \S~\ref{subsec:asymmetry} above, there is already evidence that the polarized intensity is higher on the near side than on the far side in some disks, especially HH80-81. For these sources, the polarization is more likely dominated by scattering than by either magnetically or radiatively aligned grains. This tentative conclusion needs to be strengthened (or refuted) through higher resolution observations of youngest disks (where the dust grains have grown, so that scattering is efficient, but have yet to settle to the midplane), especially at a range of wavelengths where the disk goes from being optically thin to optically thick, with the corresponding increase in the expected degree of near-far side asymmetry.

\section{Conclusion}
\label{sec:conclusion}


%
%

Observational progress in disk polarization at (sub)millimeter wavelengths has been rapid in the last few years and is expected to accelerate in the ALMA era. However, the origin of the observed  polarization remains uncertain. Part of the reason is that (sub)millimeter polarization mechanisms such as dust scattering only became widely appreciated very recently and exploration of their basic properties is still at an early stage. In this paper, we seek to quantify the effects of optical depth on the scattering-induced polarization through a combination of analytical illustration, approximate semi-analytical modeling using formal solution to the radiative transfer equation, and Monte Carlo simulations, and to evaluate their potential for probing vertical dust settling in the disk and for distinguishing the polarization from scattering from that emitted by aligned grains. The main results are summarized as follows.

\begin{itemize}

\item 
Our analytic, 1D (plane-parallel) slab model demonstrates that scattering can produce a detectable level of polarization, along the direction of the normal to the slab surface projected onto the plane of the sky (i.e., the direction of the ``local minor axis" if the slab represents a patch of the disk surface; see Fig.~\ref{fig:cartoon}), even when the optical thickness goes to infinity. The degree of polarization from such a semi-infinite slab increases with the inclination angle ($i$) of the slab with respect to the line of sight ($i=0$ means face-on view) until the slab  is viewed nearly edge-on (see Fig.~\ref{fig:pincl}). For a given inclination angle, as the total optical depth of the slab $\tau_{\rm max}$ increases from zero to infinity, the degree of polarization first increases from zero to a maximum value near $\tau_{\rm max}\sim 1$, before asymptoting to a finite value (see Fig.~\ref{fig:ptau}). This behavior is very different from that of the aligned grain case, where the polarization degree decreases monotonically with the optical depth $\tau_{\rm max}$, asymptoting to zero for a semi-infinite, isothermal slab. In addition, in optically thick regions, the degree and orientation of the polarization from direct emission by aligned grains depend strongly on the temperature distribution.
%
%

\item 
In an optically thick disk, the observed light comes mainly from the disk surface facing the observer. The shape of that surface has a strong imprint on both the degree and orientation of the scattering-induced polarization in an inclined disk. Specifically, if the scattering dust grains are vertically distributed with a significant angular thickness, the near side of the disk surface would be viewed more edge-on than the far-side (see Fig.~\ref{fig:loc_incl}), leading to a higher polarized intensity on the near-side than the far-side, especially along the minor axis. Another consequence of the finite (dust) thickness is that the polarization orientations close to the center are no longer parallel to the minor axis, which is a hallmark of scattering-induced polarization in an optically and/or geometrically thin disk. The deviation is especially clear on the major axis, where the polarization orientations rotate away from the minor axis in one direction on one side of the disk and in the opposite direction on the other side (see Fig.~\ref{fig:image}, right column). The near-far side asymmetry in polarized intensity and bifurcation in polarization orientation are quantified through an approximate semi-analytic solution to the radiative transfer equation under the single scattering approximation (see Fig.~\ref{fig:image}) and Monte Carlo simulations that include multiple scattering (see Fig.~\ref{fig:diffa}, right column).

\item
Both the near-far side asymmetry in polarized intensity and bifurcation in polarization orientation are unique to the scattering-induced polarization in an optically and geometrically thick (dust) disk. They are produced by simple geometric effects that are not shared by other mechanisms such as direct emission from (non-spherical) grains aligned with either magnetic fields or the direction of radiative flux. As such, they are robust signatures of the scattering-induced polarization, provided that the (dust) disk is both optically and geometrically thick. Both of these signatures disappear for an optically and/or geometrically thin disk.

\item 
We find anecdotal evidence from high-resolution (unpolarized) dust continuum imaging of edge-on disks that large grains are not yet settled in the youngest, Class 0 disks, but become more so in older (Class I and II) disks (\S~\ref{subsec:edge-on}). This trend is corroborated by the polarization data in inclined disks that, although still rather limited, appear to indicate that younger sources tend to have brighter polarized emission on the near-side than the far-side and thus less grain settling if scattering is responsible for the polarization (\S~\ref{subsec:asymmetry}). If confirmed, the trend would have far-reaching implications for disk grain evolution, which lies at the heart of the formation of planetesimals and ultimately planets.
\end{itemize}

ACKNOWLEDGEMENT

HY is supported in part by an SOS award from NRAO. This work is supported in part by NASA NNX 14AB38G and NSF AST-1313083 and 1716259. J.M.G. acknowledges support from MICINN AYA2014-57369-C3-P (Spain).

\end{document}